\documentclass[longauth]{aa}  
\usepackage{ulem}

\usepackage{graphicx}
\usepackage{subfigure}

\usepackage{txfonts,textcomp}
\usepackage[switch,modulo]{lineno}

\usepackage[colorlinks=true,linkcolor=red,anchorcolor=blue,citecolor=blue,filecolor=black,urlcolor=blue]{hyperref}

\usepackage[dvipsnames]{xcolor}
\usepackage{orcidlink}

\begin{document}

    \title{
Spatially resolved H$\alpha$ emission in B14-65666: Compact starbursts, ionizing efficiency, and gas kinematics in an advanced merger at the Epoch of Reionization}

   \author{C. Prieto-Jim\'enez\inst{\ref{inst:CAB}}$^{,}$ \inst{\ref{inst:UCM}} \orcidlink{0009-0005-4109-161X}, J. \'Alvarez-M\'arquez\inst{\ref{inst:CAB}}\orcidlink{0000-0002-7093-1877},
   L. Colina \inst{\ref{inst:CAB}}\orcidlink{0000-0002-9090-4227},  A. Crespo Gómez \inst{\ref{inst:CAB}}$^{,}$ \inst{\ref{inst:STScI}}\orcidlink{0000-0003-2119-277X},  A. Bik \inst{\ref{inst:Stockholm}}\orcidlink{0000-0001-8068-0891}, G. Östlin \inst{\ref{inst:Stockholm}}\orcidlink{0000-0002-3005-1349}, A.~Alonso-Herrero \inst{\ref{inst:CAB_ESAC}}\orcidlink{0000-0001-6794-2519}, L. Boogaard\inst{\ref{inst:leiden}}\orcidlink{0000-0002-3952-8588}, K. I. Caputi \inst{\ref{inst:Kapteyn}}$^{,}$ \inst{\ref{inst:DAWN}}\orcidlink{0000-0001-8183-1460}, L. Costantin \inst{\ref{inst:CAB}}\orcidlink{0000-0001-6820-0015}, A. Eckart \inst{\ref{inst:koln}}\orcidlink{0000-0001-6049-3132}, M.~García-Marín \inst{\ref{inst:ESA_STSCI}}\orcidlink{0000-0003-4801-0489}, S.~Gillman \inst{\ref{inst:DAWN}}$^{,}$ \inst{\ref{inst:DTU}}\orcidlink{0000-0001-9885-4589}, J. Hjorth \inst{\ref{inst:DARK}}\orcidlink{0000-0002-4571-2306}, E. Iani \inst{\ref{inst:Kapteyn}}\orcidlink{0000-0001-8386-3546}, I. Jermann \inst{\ref{inst:DAWN}}$^{,}$ \inst{\ref{inst:DTU}}\orcidlink{0000-0002-2624-1641}, A. Labiano \inst{\ref{inst:Telespazio}}\orcidlink{0000-0002-0690-8824}, D. Langeroodi \inst{\ref{inst:DARK}}\orcidlink{0000-0001-5710-8395}, J. Melinder \inst{\ref{inst:Stockholm}}\orcidlink{0000-0003-0470-8754}, T.~ Moutard \inst{\ref{inst:ESA_ESAC}}\orcidlink{0000-0002-3305-9901}, F. Peißker \inst{\ref{inst:koln}}\orcidlink{0000-0002-9850-2708}, P. G. Pérez-González  \inst{\ref{inst:CAB}}\orcidlink{0000-0003-4528-5639}, J. P. Pye \inst{\ref{inst:leicester}}\orcidlink{0000-0002-0932-4330}, P. Rinaldi \inst{\ref{inst:tucson}}\orcidlink{0000-0002-5104-8245}, T. V. Tikkanen \inst{\ref{inst:leicester}}\orcidlink{0009-0003-6128-2347}, P.~van~der~Werf~\inst{\ref{inst:leiden}}~\orcidlink{0000-0001-5434-5942}, F. Walter \inst{\ref{inst:MPIA}}\orcidlink{0000-0003-4793-7880}, T. Hashimoto \inst{\ref{inst:Tsukuba}}$^{,}$ \inst{\ref{inst:Tomonaga}}\orcidlink{0000-0002-0898-4038}, Y. Sugahara \inst{\ref{inst:National_Astronomical}}$^{,}$ \inst{\ref{inst:Waseda}}\orcidlink{0000-0001-6958-7856}, M. G\"udel \inst{\ref{inst:ETH}}\orcidlink{0000-0001-9818-0588}, T. Henning \inst{\ref{inst:MPIA}}\orcidlink{0000-0002-1493-300X}}

    \institute{Centro de Astrobiolog\'{\i}a (CAB), CSIC-INTA, Ctra. de Ajalvir km 4, Torrej\'on de Ardoz, E-28850, Madrid, Spain\\  \email{cprieto@cab.inta-csic.es} \label{inst:CAB}
    \and Departamento de F\'{i}sica de la Tierra y Astrof\'{i}sica,
    Facultad de Ciencias F\'{i}sicas, Universidad Complutense de Madrid, E-28040, Madrid, Spain \label{inst:UCM} \and Space Telescope Science Institute (STScI), 3700 San Martin Drive, Baltimore, MD 21218, USA \label{inst:STScI} \and Department of Astronomy, Stockholm University, Oscar Klein Centre, AlbaNova University Centre, 106 91 Stockholm, Sweden \label{inst:Stockholm} \and Centro de Astrobiolog\'{\i}a (CAB), CSIC-INTA, Camino Bajo del Castillo s/n, E-28692 Villanueva de la Ca\~nada, Madrid, Spain \label{inst:CAB_ESAC} \and Leiden Observatory, Leiden University, PO Box 9513, 2300 RA Leiden, The Netherlands \label{inst:leiden} \and Kapteyn Astronomical Institute, University of Groningen, P.O. Box 800, 9700AV Groningen, The Netherlands \label{inst:Kapteyn} \and Cosmic Dawn Center (DAWN), Copenhagen, Denmark \label{inst:DAWN} \and Physikalisches Institut der Universität zu Köln, Zülpicher Str. 77, 50937 Köln, Germany \label{inst:koln}
    \and European Space Agency, Space Telescope Science Institute, Baltimore, Maryland, USA \label{inst:ESA_STSCI}\and DTU-Space, Elektrovej, Building 328, 2800, Kgs. Lyngby, Denmark \label{inst:DTU} \and DARK, Niels Bohr Institute, University of Copenhagen, 2200 Copenhagen, Denmark \label{inst:DARK} \and Telespazio UK for the European Space Agency (ESA), ESAC,  Camino Bajo del Castillo s/n, 28692 Villanueva de la Cañada, Madrid, Spain \label{inst:Telespazio} \and European Space Agency (ESA), European Space Astronomy Centre (ESAC), Camino Bajo del Castillo s/n, 28692 Villanueva de la Cañada, Madrid, Spain \label{inst:ESA_ESAC} \and School of Physics \& Astronomy, Space Park Leicester, University of Leicester, 92 Corporation Road, Leicester LE4 5SP, UK \label{inst:leicester}\and Steward Observatory, University of Arizona, 933 North Cherry Avenue, Tucson, AZ 85721, USA \label{inst:tucson} \and Max Planck Institut f\"ur Astronomie, K\"onigstuhl 17, D-69117, Heidelberg, Germany \label{inst:MPIA}
    \and Division of Physics, Faculty of Pure and Applied Sciences, University of Tsukuba,Tsukuba, Ibaraki 305-8571, Japan \label{inst:Tsukuba}
    \and Tomonaga Center for the History of the Universe (TCHoU), Faculty of Pure and Applied Sciences, University of Tsukuba, Tsukuba, Ibaraki 305-8571, Japan \label{inst:Tomonaga}
    \and National Astronomical Observatory of Japan, 2-21-1 Osawa, Mitaka, Tokyo 181-8588, Japan \label{inst:National_Astronomical}
    \and Waseda Research Institute for Science and Engineering, Faculty of Science and Engineering, Waseda University, 3-4-1 Okubo, Shinjuku, Tokyo 169-8555, Japan \label{inst:Waseda} \and ETH Z\"urich, Institute for Particle Physics and Astrophysics, Wolfgang-Pauli-Str. 27, 8093 Z\"urich, Switzerland \label{inst:ETH}
   }

   \date{Received 6 April, 2025 / Accepted 8 July, 2025}

\abstract 
{ We present MIRI/JWST medium-resolution spectroscopy (MRS) and imaging (MIRIM) of B14-65666, a source identified as a Lyman-break and interacting galaxy at a redshift of $z$\,=\,7.15. We detect the H$\alpha$ line emission in this system, revealing a spatially resolved structure of the H$\alpha$-emitting gas, which consists of two distinct galaxies, E and W, at a projected distance of 0.4 arcsec apart (i.e., 2.2 kpc).
One of the galaxies (E) is very compact (upper limit for the effective radius of 63 pc) in the rest-frame ultraviolet light, while the other galaxy (W) is more extended (effective radius of 348 pc), showing a clumpy structure reminiscent of a tidal tail.
The total H$\alpha$ luminosity implies that the system is forming stars at a rate of 76\,±\,8 M$_{\odot}$\,yr$^{-1}$ and 30\,±\,4 M$_{\odot}$\,yr$^{-1}$ for E and W galaxies, respectively. The ionizing photon production efficiency, $\log(\zeta_\mathrm{ion}$), for galaxies E and W, has values of 25.1\,±\,0.1 Hz erg$^{-1}$ and 25.5\,±\,0.1 Hz erg$^{-1}$, which is within the range measured in galaxies at similar redshifts. The high values derived for the H$\alpha$ equivalent widths (832\,±\,100 and 536\,±\,78 $\AA$) and the distinct locations of the E and W galaxies in the $\log(\zeta_\mathrm{ion}$) $-$ equivalent width (H$\alpha$) plane indicate that the system is dominated by a young (under 10 Myr) stellar population.  The overall spectral-energy distribution suggests that in addition to a young stellar population, the two galaxies may have mature (over 100 Myr) stellar populations and very different dust attenuations, with galaxy E showing a larger attenuation (A$_{V}$\,=\,1.5 mag) compared to the almost dust-free (A$_{V}$\,=\,0.1 mag) galaxy W. The derived star formation rate (SFR) and stellar masses identify the two galaxies as going through a starburst phase characterized by a specific SFR (sSFR) of 40\,--\,50\,Gyr$^{-1}$. Galaxy E has an extreme stellar mass surface density (6\,$\times$\,10$^4$\,M$_{\odot}$\,pc$^{-2}$), close to that of the nuclei of low-$z$ galaxies, while galaxy W (10$^3$\,M$_{\odot}$ pc$^{-2}$) is consistent with the surface densities measured in galaxies at these redshifts. 
The kinematics of the ionized gas traced by the H$\alpha$ line show a velocity difference of 175\,$\pm$\,28 km\,s$^{-1}$ between the two components of B14-65666 and a broader profile for galaxy W (312\,$\pm\,$44 km\,s$^{-1}$) relative to galaxy E (243\,$\pm$\,41 km\,s$^{-1}$). 
The detailed study of B14-65666 shows that the complex stellar and interstellar medium structure in merging galaxy systems was already in place by the Epoch of Reionization. The general properties of B14-65666 agree with those predicted for massive merging systems at redshifts of 7 and above in the FIRSTLIGHT cosmological simulations. 
The in-depth study of systems such as B14-65666 reveal how galaxy mergers in the early Universe drive intense star formation, shape the interstellar medium, and influence the buildup of stellar mass, just 700\,--\,800 Myr after the Big Bang.
}

\titlerunning{MIRI analysis of B14-65666}
\authorrunning{C. Prieto-Jiménez }

\maketitle

\section{Introduction}\label{Sect:intro}

The \textit{James Webb Space Telescope} (JWST, \citealt{Gardner+23}) is transforming our understanding of galaxy formation, particularly during the so-called Epoch of Reionization (EoR). The JWST has photometrically detected potential galaxy candidates up to redshift 25 (\citealt{perezgonzalez2025risegalacticempireluminosity}) and 
spectroscopically confirmed galaxies up to redshift 14.44 (\citealt{naidu_2025}). At these redshifts, the rest-frame optical emission lines and continuum are redshifted into the mid-infrared (5\,--\,25\,$\mu$m), which is the spectral range covered by the Mid-Infrared Instrument (MIRI, \citealt{Rieke+15,Wright+15,Wright+23}). At redshifts of 7 and above, this instrument allows us to explore the strong optical emission lines, such as H$\alpha$.

Mergers are expected to have a greater impact on galaxy evolution at higher redshifts than at lower redshifts, as the merger rate increases with redshift (\citealt{duan_2024_mergers}). The gravitational forces from mergers can cause significant disruptions in galaxy structures and movements, leading to features such as tidal tails  and asymmetric stellar envelopes (e.g., \citealt{toomre72}; \citealt{mihos_1996}; \citealt{Bushouse_2002}; 
\citealt{conselice_2003};
\citealt{Spilker+22}). These disturbances can cause gas to flow from the outer regions into the center of galaxies, triggering the formation of new stars in central areas (\citealt{perez-gonzalez_24_jekyll_hyde}). Simulations such as FirstLight (\citealt{Ceverino_17}) show that about 10\% of galaxies at redshifts of 6\,--\,10 exhibit clumpy structures, which are primarily formed through mergers. The central clumps are mainly composed of older stellar populations exceeding 50\,Myr, while off-center clumps originate from gas debris along tidal tails during major mergers. These mergers have a short life, typically merging into the main galaxy within a few tens of million years  (\citealt{nazakoto_2024}). Observational studies also support the presence of mergers in the EoR, with recent works identifying clear merger signatures in high-redshift galaxies (e.g., \citealt{Alvarez-Marquez+23-SPT}; \citealt{arribas_24_spt};  \citealt{harshan_2024}; \citealt{hu_papovich_2025}).

High-$z$ star-forming galaxies (SFGs) are characterized by optical lines with high (rest-frame) values of the equivalent widths (EW up to 3000\,$\AA$) in extreme emission line galaxies (\citealt{boyett24}). There is also a trend of EW increasing with redshift, which indicates an evolution in the galaxy-specific star formation rate (\citealt{Rinaldi+23}; \citealt{rinaldi2024emergencestarformationmain}) and young stellar populations (<\,10\,Myr). Studies at low redshifts have shown that galaxies in close pairs exhibit an SFR increase compared to isolated galaxies, highlighting the impact of gravitational interactions on gas dynamics and star formation (\citealt{barton_2000}). This enhancement is also observed at higher redshifts (6.5\,--\,8.5), with an excess in SFR of 0.26\,±\,0.11 dex found in very close pairs, with no significant increase at larger separations (\citealt{duan_2024}; \citealt{puskas25}).

B14-65666 (also known as the Big Three Dragons galaxy) was discovered based on wide-field imaging data of the UltraVISTA survey of the Cosmological
Evolution Survey (COSMOS) field in Y, J, H, and K${_s}$ near-infrared bands (\citealt{bowler_2012}, \citealt{bowler_2014_the}). It was then spectroscopically detected in Ly$\alpha$ (\citealt{furusawa_2016_a}). An image captured with the Hubble Space Telescope (HST) showed that the galaxy consists of two clearly separated components in the rest-frame ultraviolet (UV), thus suggesting that the object could be undergoing a merger of two galaxies (\citealt{bowler_2017_unveiling}). Observations performed with the Atacama Large Millimeter/submillimeter Array (ALMA) with Band 6 reported the detection of an $\approx$\,160\,$\mu$m dust continuum at the position of the components (\citealt{bowler_2018_obscured}). It also detected [O\,III]\,88\,$\mu$m, [C\,II]\,158\,$\mu$m, and underlying dust continuum emission at $\approx$\,90\,$\mu$m in Bands 6 and 8 (\citealt{hashimoto_2019_big}). Furthermore, they showed that [O\,III]\,88\,$\mu$m and [C\,II]\,158\,$\mu$m can be decomposed into two components that are spatially resolved and kinematically separated. These components are associated with the two UV emission peaks identified in the HST images. 
Additional observations with ALMA Band 7 did not detect the presence of the far-infrared [N\,II]\,122\,$\mu$m fine-structure  emission line, nor did ALMA Band 3 observations detect other lines such as CO (6-5), CO (7-6), and [C\,I] (2-1) (\citealt{sugahara_2021_big}; \citealt{hashimoto_2023_big}).

Recent JWST Near Infrared Camera (NIRCam) imaging has revealed the complex morphology of the two galaxy components: galaxy E consists of a compact core, surrounded by diffuse, extended, rest-frame optical emission, and galaxy W has an elongated morphology interpreted as tidal tails (\citealt{sugahara_2024_rioja}). Follow-up observations with NIRSpec IFU from 2.9\,--\,5.3\,$\mu$m directly map the strength of the rest-frame optical emission lines across the source (\citealt{gareth_2024}). The profile of these lines revealed the presence of both narrow and broad line emission, with the latter indicating tidal interaction or outflows. This analysis showed that both galaxies have subsolar metallicities (0.24\,Z$_\odot$ and 0.3\,Z$_\odot$, E and W, respectively) and a high rate of star formation. The observations do not show evidence for the presence of an active galactic nucleus (AGN) in B14-65666.

B14-65666 is one of the few galaxies identified as undergoing a major merger during the EoR. This system allows the investigation of the physical properties of an early major merger, including its stellar structure, as well as the different phases of the interstellar medium (ISM). Additionally, this system provides a reference for direct comparisons with simulations.
In this paper, we present the first spatially resolved spectroscopy of the H$\alpha$ emission in the system B14-65666 at $z$\,=\,7.15 based on new MIRI/JWST integral field spectroscopy. The paper presents several of the physical properties of B14-65666 based on the combination of the new MIRI data with ancillary NIRCam imaging and ALMA spectroscopy. 
This paper is organized as follows. In Sect. \ref{Sec:obs_cal}, we present the observations and data calibrations. In Sect. \ref{Sect:analysis}, we outline our analysis methods of spectroscopy and photometry. In Sect. \ref{Sec:results_dis}, we present the results of the analysis. In the subsections, we study the spectral energy distribution of the galaxy (\ref{subSect:SED}), the attenuation (\ref{subSec:attenuation}), the star formation rate and burstiness (\ref{subSec:SFR}), the H$\alpha$ EW and ionizing photon production efficiency (\ref{subSec:ew}), the kinematics (\ref{subSect:kinematics}), the stellar and dynamical mass (\ref{subSect:morph}), the mass-size relation (\ref{subSect:mass-size}), the starburst nature (\ref{subSect:burstiness}), and finally the origin and evolution of the system according to simulations (\ref{subSect:simulations}). We discuss and analyze the spectral and spatial distribution of the H$\alpha$ line and the derived properties of the galaxies, and we study the morphology and SFR of the galaxy and compare it with local galaxies. Our conclusions are presented in Sect. \ref{Sec:conclusion_Summary}.

In this paper, we assume a flat $\Lambda$CDM cosmology with $\Omega_\mathrm{m}$\,=\,0.310, and $H_0$\,=\,67.7\,km\,s$^{-1}$\,Mpc$^{-1}$ \citep{PlanckCollaboration18VI}. With these parameters, 1\," is equal to 5.275 kpc at $z$\,=\,7.15 and the luminosity distance is $D_L$=\,72276.9 Mpc. For these cosmological parameters and redshift, the age of the Universe corresponds to 740 Myr. We use vacuum emission-line wavelengths throughout the paper.

\section{Observations and data calibrations}\label{Sec:obs_cal}

\subsection{MIRI}

B14-65666 was observed with MIRI on 30 April, 2023 as a part of the MIRI European Consortium guaranteed time observations (program ID 1284, PI: L. Colina). The observations are composed of MIRI imaging (MIRIM, \citealt{Bouchet+15, dicken_2024_miri}) and integral field spectroscopy using the Medium Resolution Spectrograph (MRS; \citealt{Wells+15,Argyriou+23}).

\subsubsection{MIRI spectroscopy}

The MRS observations were performed with the SHORT band for channels 1\,--\,4. We only used Channel 1 in this study. This channel covers the wavelengths 4.9\,--\,5.74\,$\mu$m, thus targeting the H$\alpha$ emission line at redshift 7.15 and other fainter emission lines such as [N\,II]\,6550,6585\,$\AA$ and [S\,II]\,6718,6733\,$\AA$. The total on-source exposure time of the MRS observations is 22743  seconds. The positive and negative four-point dither patterns for point sources were used. For each dither position, a total of six integrations were obtained using the SLOWR1 readout mode with 19 groups each. When the MRS targeted B14-65666, a parallel field was imaged using the MIRI F770W filter. This imaging data is utilized exclusively for the astrometric calibration of the MRS (see Sect. \ref{Sect:astrometry}).

We used version 1.15.1 of the JWST calibration pipeline and context 1276 of the Calibration Reference Data System (CRDS) to process and calibrate the MRS observations of B14-65666. The standard MRS pipeline procedure (\citealt{bushouse_pipeline}) was used with additional customized steps to improve the quality of the final MRS-calibrated products and background (see \citealt{Alvarez-Marquez+23-SPT} and \citealt{Alvarez-Marquez+23-SPT} for details). Additionally, we implemented the algorithm developed by \cite{spilker_2023_spatial} to correct the striping pattern produced by the cosmic-ray showers in the spectral cube. The final Channel 1 cube has a spatial and spectral sampling of 0.13"\,$\times$\,0.13"\,$\times$\,0.8\,nm (\citealt{Law+23}), and the resolving power ($R\,\equiv\,\lambda/\Delta\lambda$) is about 3500 (\citealt{Labiano+21}; \citealt{Jones+23}).

\subsubsection{MIRI imaging} \label{subsect:miri_2}

The MIRIM observations of B14-65666 were performed using the F560W filter covering the rest-frame optical continuum at $\approx$\,0.7\,$\mu$m and the H$\alpha$ emission line. The MIRIM F560W image has a total on-source integration time of 2675 seconds. The observational setup consists of a four-point dither pattern, with the dither composed of two integrations of 120 groups using the FASTR1 readout mode.

The MIRIM F560W observations are calibrated with version 1.12.0 of the JWST pipeline  \citep{bushouse_pipeline} and context 1150 of the CRDS. We followed the general procedure to calibrate the data, but we also included more steps to correct for striping artifacts and background gradients (see \citealt{pablo_perezgonzalez_2024_LRD}; \citealt{goran_ostlin_2024_midis} for details). The final MIRIM F560W mosaic has a pixel size of 0.06\,", after resampling, and a field of view (FoV) of  74"\,$\times$\,113\,". The point spread function (PSF) full width at half maximum (FWHM) is 0.207\," (\citealt{libralato_2024_highprecision}).

\subsection{NIRCam} \label{subsect:nircam_2}
The NIRCam \citep{Rieke_2023} data used in this work were presented in a previous study (\citealt{sugahara_2024_rioja}). The observations were taken with filters F115W, F150W, F200W, F277W, F356W, and F444W (JWST GO1 program ID 1840; PIs: J. Álvarez-Márquez and T. Hashimoto). At redshift 7.15, F115W, F150W, and F200W sample the rest-frame UV spectrum, F277W and F356W cover the Balmer break region, and F444W targets the rest-frame continuum around 5000\,\(\text{\AA}\) and the [O\,III]\,4960,5008\,$\AA$ emission lines.

NIRCam and MIRIM images present large variations on the spatial resolution, with PSF FWHMs increasing with wavelength from 0.04\," to 0.21\,". To obtain consistent multiwavelength aperture photometry in the different NIRCam and MIRIM images, we performed a PSF homogenization in the NIRCam images to match the spatial resolution of NIRCam images to the coarser MIRIM F560W image. However, we also used the NIRCam images with the native spatial resolution to investigate the structure of B14-65666 in the rest-frame UV (see Sect. \ref{subSect:reff}).

The PSF homogenization procedure includes the following steps. We employed the observed effective PSFs (ePSFs) for MIRI (\citealt{libralato_2024_highprecision}) and for NIRCam  (\citealt{nardiello_2022})\footnote{\url{https://www.stsci.edu/~jayander/JWST1PASS/LIB/PSFs/STDPSFs/}.}. We compared these ePSFs with point-like sources from our observations and conclude that they are consistent with each other. The first step was to rotate the NIRCam ePSF to the position angle (PA) of the MIRI image (PA=108º), ensuring orientation alignment between the two ePSFs. We rotate the MIRI ePSF to match the PA of the MIRI image. Next, we resampled the pixel scale of the MIRI ePSF (0.028\,", \citealt{libralato_2024_highprecision}) to match that of the MIRI imager and the NIRCam ePSF (0.008\," for F115W, F150W, and F200W; and 0.016\," for F277W, F356W, and F444W; \citealt{nardiello_2022}) to match the pixel scale of the MIRI imager. Using these aligned and resampled ePSFs, we then generated a matching kernel using the python package \texttt{PHOTUTILS} to bridge the differences between the NIRCam and MIRI ePSFs. Finally, to apply the kernel using the python function \texttt{creatematchingkernel} (\citealt{larry_bradley_2022_6825092}), we rescaled it to the NIRCam image pixel size (0.03\," after resampling). Subsequently, we convolved the NIRCam images with the kernel, thus homogenizing the PSF. 

\subsection{ALMA ancillary data} 

The ALMA data used in this work were previously presented in other studies. ALMA Band-6 and Band-8 data were taken in Cycles 4 (ID 2016.1.00954.S, PI: A. K. Inoue) and 5 (ID 2017.1.00190.S, PI: A. K. Inoue) to target the [C\,II]\,158\,$\mu$m and [O\,III]\,88\,$\mu$m emission lines (\citealt{hashimoto_2019_big}). ALMA Band-7 data were taken during Cycle 7 (ID 2019.1.01491.S, PI: A. K. Inoue; \citealt{sugahara_2021_big}). In this article, we used the spectra of [C\,II]\,158\,$\mu$m and [O\,III]\,88\,$\mu$m and the emission-line and continuum (90, 120, and 160\,$\mu$m) fluxes presented in \cite{hashimoto_2019_big} and  \cite{sugahara_2021_big}.

\subsection{MIRI-NIRCam astrometry}\label{Sect:astrometry}
We corrected the astrometry in the MIRI F560W image by using the centroid of two point-like sources within the FoV of the available GAIA DR3 \citep{GaiaCollaboration+22} coordinates. The uncertainty in the final positioning is 18 mas. With the corrected MIRIM F560W image, we aligned the simultaneous MIRIM F770W image with two point-like sources in the common FoV of both images. As the MRS data and the MIRIM F770W image were taken simultaneously, we used the astrometric offset found in the MIRIM F770W image to correct the MRS 1SHORT cube, obtaining an uncertainty of less than 60\,mas (0.3\,kpc). Finally, we aligned the NIRCam images using the astrometrically corrected MIRIM F560W image as we did for the simultaneous MIRIM F770W image. The NIRCam images have an uncertainty smaller than 35\,mas; this was obtained with the available star and galaxy that have Gaia astrometry \citep{GaiaCollaboration+22}.

\section{Analysis}\label{Sect:analysis}

\begin{figure*} 
    \centering   \includegraphics[width=0.90\linewidth]{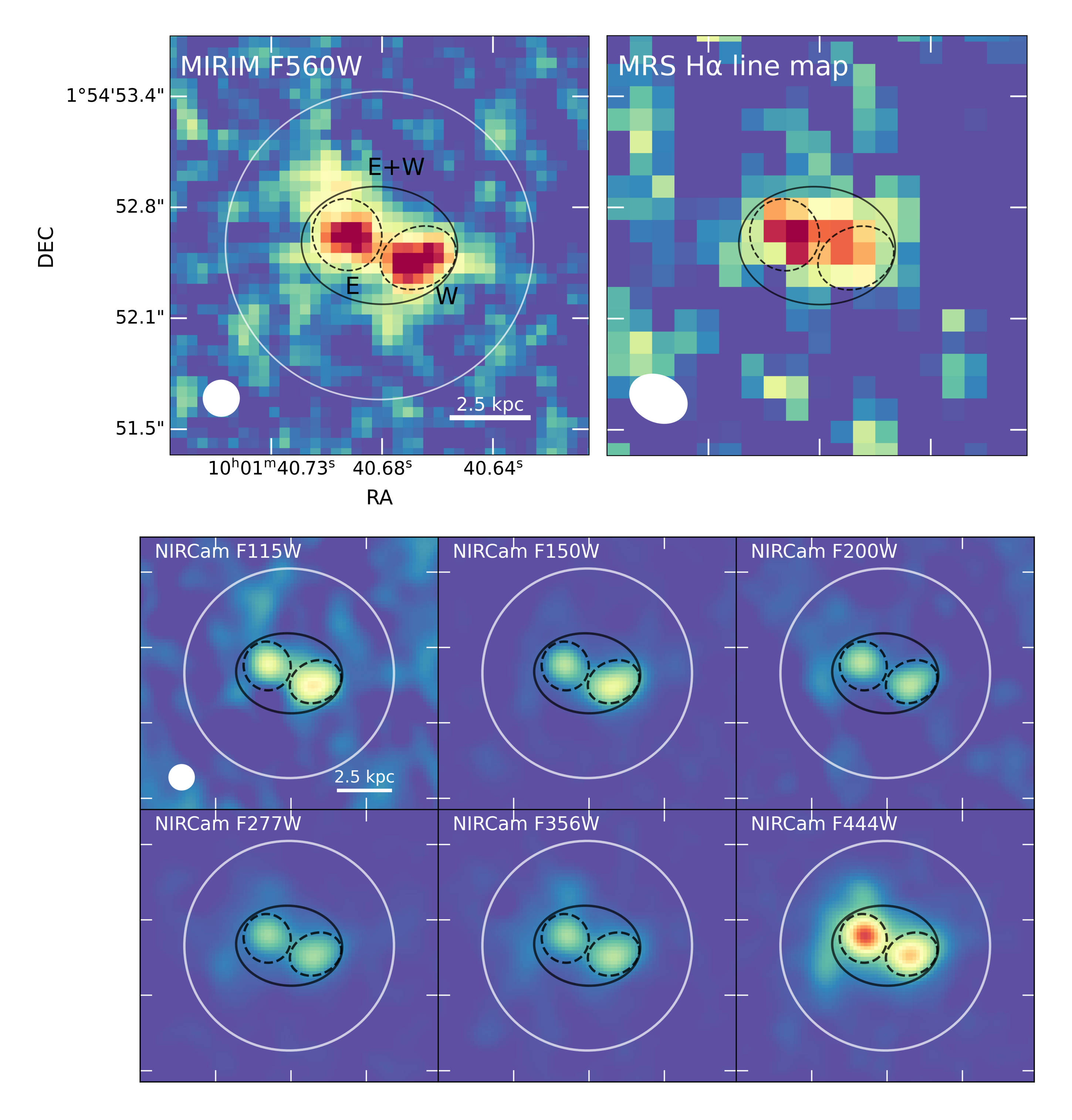}
    \caption{ \textit{Top left}: MIRI Image F560W. The white circumference represents a 0.9\," radius aperture. The solid black elliptical line represents the E+W galaxies' apertures for the H$\alpha$ emission of the whole galaxy; the dashed black lines represent the apertures for galaxies E and W. The white filled circle area in the bottom left represents the spatial resolution (PSF FWHM) of the MIRI image F560W. \textit{Top right}: H$\alpha$ line map. The H$\alpha$ line map is generated by integrating H$\alpha$ line emission in the velocity range: -500\,<\,{\it v}\,[km\,s$^{-1}$]\,<\,300. The white filled area represents the spatial resolution (PSF FWHM) of the MRS. \textit{Bottom}: NIRCam cutouts in six filters, F115W to F444W (left to right and top to bottom). The images are homogenized to the PSF of the MIRI image, represented with the white filled circle in the NIRCam F115W image.}
    \label{fig:all}
\end{figure*}
\subsection{MIRI and NIRCam images and H$\alpha$ line map} \label{subsect:MIRI_NIRCAM_IMAGES_MAP}

Previous HST images showed that B14-65666 is composed of two spatially separated galaxies in the rest-frame UV, a brighter galaxy in the northeast (galaxy E), and a fainter galaxy in the southwest (galaxy W) (\citealt{bowler_2017_unveiling}). ALMA data further demonstrated that the [O\,III]\,88\,$\mu$m and [C\,II]\,158\,$\mu$m emission lines can be spatially and spectrally decomposed into two components associated with galaxies E and W (\citealt{Hashimoto+18}). NIRCam images revealed the complex morphology of the individual galaxy components (\citealt{sugahara_2024_rioja}): galaxy E has a compact core surrounded by diffuse extended optical emission, while galaxy W displays elongated clumpy morphology. 

The MIRIM and MRS data cover the emission line of H$\alpha$ together with the continuum at 0.7\,$\mu$m. MIRI observations show a similar trend to NIRCam images. MRS shows an extended H$\alpha$ emission associated with E and W galaxy components; additionally, each of these galaxy components is spectroscopically resolved as they were already presented in ALMA  observations (\citealt{hashimoto_2019_big}). We present MIRI and NIRCam images in Fig. \ref{fig:all}.

For this study, we defined a total elliptical aperture for the whole system (named galaxies E+W) with a semi-major axis of 0.456\,", semi-minor axis of 0.343\,", and position angle (PA) of 85.83º.  
In the case of the galaxy E we define an elliptical aperture with semi-major axis of 0.211\," and semi-minor axis of 0.201\,", position angle (PA) of 112.83º. 
For the galaxy W, we extracted an ellipse with a 0.228\," and 0.177\," semi-major and semi-minor axis, respectively, and a PA of 112.83º. The coordinates of the center are shown in Table \ref{tab:results}. 
The E and W elliptical apertures are compatible with the main emission of these galaxy components in our MIRIM F560W image. The elliptical
aperture of galaxies E+W covers the extended H$\alpha$ emission from the MRS H$\alpha$ line map. The H$\alpha$ line map is generated by integrating the H$\alpha$ line emission, centered at the H$\alpha$ peak of galaxies E+W (5.3510\,$\mu$m) in the velocity range -500\,<\,{\it v}\,[km\,s$^{-1}$]\,<\,300 (see Fig. \ref{fig:all}).

\begin{figure*}
\centering
\includegraphics[width=0.97\textwidth]{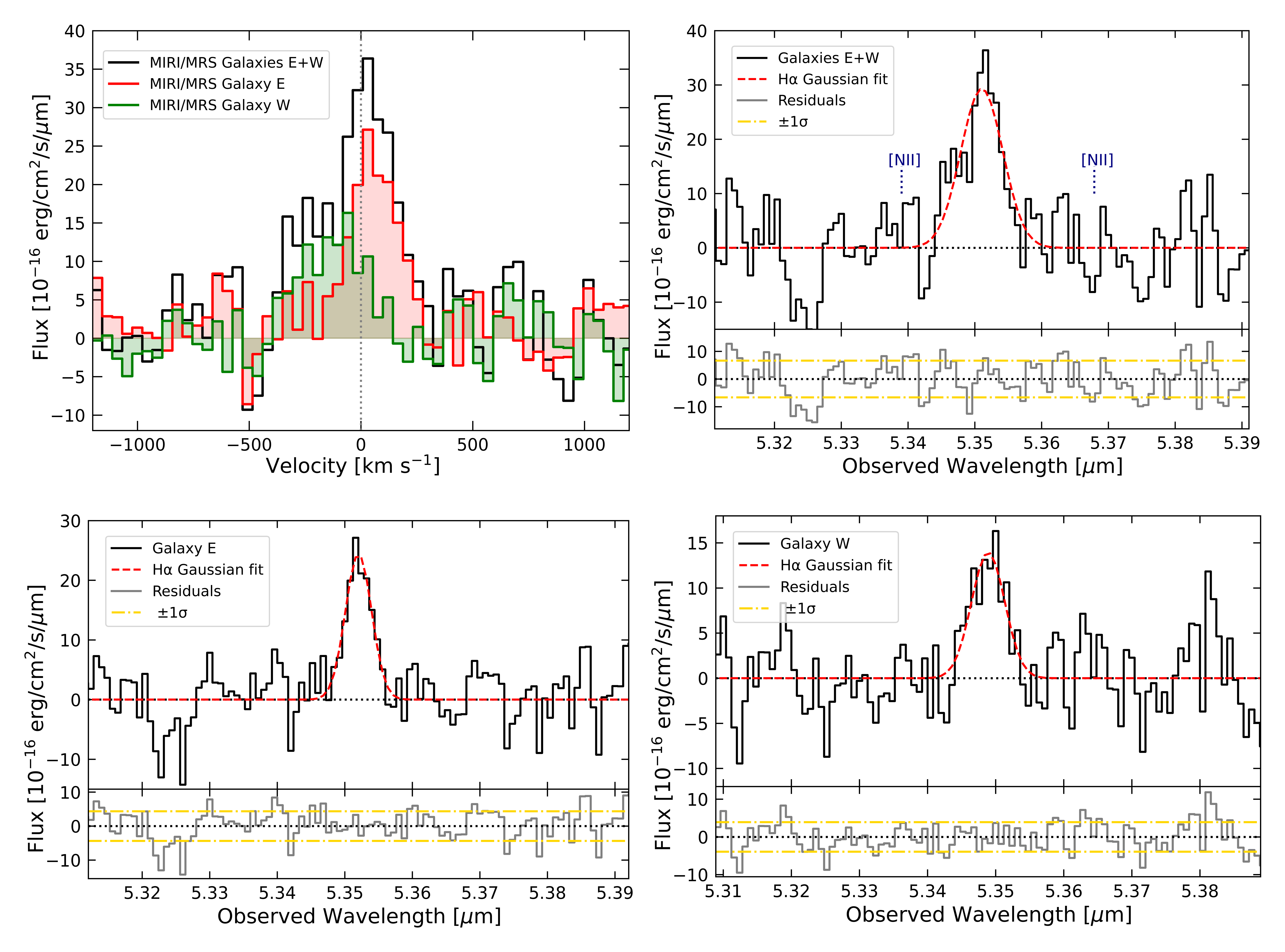}
\caption{MIRI MRS 1 SHORT spectra for H$\alpha$ observed at \textit{z}\,=\,7.1513. \textit{Top left}: H$\alpha$ spectra shown in velocity space for two spatially separated galaxies, E and W (red line and green line, respectively), identified in the H$\alpha$ line map. The black line shows the integrated MRS spectrum of B14-65666 extracted using the galaxies E+W aperture. \textit{Top right}: E+W galaxies' integrated H$\alpha$ spectrum. A vertical dashed line shows the wavelength of the weaker undetected lines [N\,II]\,6550\,\(\text{\AA}\) and\,[N\,II]\,6585\,\(\text{\AA}\) that are well separated in wavelength from  H$\alpha$\,6563\,\(\text{\AA}\). \textit{Bottom left}: Galaxy E's integrated H$\alpha$ spectrum. \textit{Bottom right}: Galaxy W's integrated H$\alpha$ spectrum. The red dashed line shows the single Gaussian fits, with the residuals shown in the bottom panels. $1\sigma$ noise level is shown in yellow; it was calculated from the continuum adjacent to the emission line.}
\label{fig:main_figure}
\end{figure*}

\begin{table*}[]

\centering
\caption{Summary of observed quantities for B14-65666.}
\begin{tabular}{lcccc}
\hline
Parameter & Galaxies E+W & Galaxy E & Galaxy W \\
\hline
RA (ICRS) & 10:01:40.6858s & 10:01:40.6984s & 10:01:40.6707s \\
DEC (ICRS) & +1:54:52.553s & +1:54:52.616s & +1:54:52.481s \\
Redshift & $7.1513 \pm 0.0007$ & $7.1529 \pm 0.0004$ & $7.1481 \pm 0.0007$ \\
FWHM [km\,s$^{-1}$] & $407 \pm 86$ & $ 243 \pm  41$ & $312 \pm 44$ \\
Flux H$\alpha$ [$\times 10^{-18}$ erg s$^{-1}$ cm$^{-2}$] & $23.1 \pm 3.1 $ & $11.9 \pm 1.3 $ & $8.6  \pm 1.2 $ \\
H$\alpha$ peak [$\mu$m] & $5.3510 \pm 0.0004$ & $ 5.3520 \pm 0.0003$ & $ 5.3489 \pm 0.0004$ \\
Luminosity H$\alpha$ [$\times 10^{42}$ erg s$^{-1}$] & $14.4 \pm 2.0 $ & $7.4 \pm 0.8 $ & $5.3  \pm 0.7 $ \\

Flux MIRIM F560W [$\mu$Jy]& $0.57\pm 0.02$ & $0.21 \pm 0.01$ & $0.22 \pm 0.01$ \\
Flux F115W [$\mu$Jy] & $ 0.222\pm  0.032$ & $ 0.073\pm 0.011 $ & $  0.114\pm 0.010 $ \\
Flux F150W [$\mu$Jy]& $ 0.313\pm 0.010$ & $  0.096\pm 0.003$ & $ 0.151\pm 0.003$ \\
Flux F200W [$\mu$Jy]& $ 0.288\pm 0.029 $ & $  0.114\pm 0.001  $ & $  0.119\pm 0.010 $ \\
Flux F277W [$\mu$Jy]& $ 0.380\pm 0.100$ & $ 0.132\pm 0.003 $ & $ 0.145\pm 0.003$ \\
Flux F356W [$\mu$Jy]& $ 0.370\pm 0.007$ & $ 0.125\pm  0.002 $ & $ 0.147\pm 0.002$ \\
Flux F444W [$\mu$Jy]& $ 0.761\pm 0.006$ & $ 0.316\pm 0.002$ & $ 0.245\pm 0.002$ \\

\hline
\end{tabular}

\tablefoot{Photometric results obtained for MIRIM and NIRCam images and spectroscopic results obtained with the MRS.}

\label{tab:results}
\end{table*}

\subsection{H$\alpha$ spectra and line fluxes}\label{subSect:halpha_spectra}

We extracted the integrated H$\alpha$ spectra of galaxies E+W, E, and W using the three defined apertures (Fig. \ref{fig:all}). We corrected the two individual spectra for aperture losses, assuming that the two galaxies are unresolved sources for the MRS angular resolution taking into account the MRS PSF (\citealt{Argyriou+23}), which has a FWHM of 0.27\,"\,$\times$\,0.35\," at 5.35\,$\mu$m. The aperture correction is 1/0.57 for the galaxies E and W. In the case of the galaxies E+W, we also corrected for aperture losses, assuming that the flux is mainly dominated by galaxies E and W, which are unresolved. The aperture correction in this case is 1/0.81. We also corrected the spectrum for the contamination of flux from galaxy E in the W aperture by modeling galaxy E at the center of the aforementioned PSF and measuring how much of its flux would appear in the W aperture. We applied the same procedure for galaxy W to estimate its contribution to the E aperture. These corrections are 1/0.02 and 1/0.03, respectively.

We extracted twelve 1D spectra using the same apertures as the E+W, E, and W galaxies, respectively, at random positions of the MRS FoV clean of emission. We combined these spectra to generate the 1D median of the local background. The median is subtracted from the H$\alpha$ spectra with the aim of removing any systematic residual feature left in the MRS calibration process. The three spectra are presented in Fig. \ref{fig:main_figure}. 

The H$\alpha$ spectrum is modeled by a single Gaussian function, to fit the emission line, and a second-order polynomial, to fit any residual background gradient. Using a Monte Carlo simulation, we estimated the uncertainties for the derived emission line parameters (such as FWHM, flux, and central wavelength). To determine the noise level in the spectrum, the root mean square (rms) was calculated from the continuum adjacent to the emission line. Using this rms value as the 1$\sigma$ standard deviation, we generated 1000 synthetic spectra by adding random Gaussian noise to the original spectrum. Each synthetic spectrum was then refit to obtain new measurements of the line parameters. The final uncertainty for each parameter is given by the standard deviation of these measurements across all simulations. The Gaussian fit is shown in Fig. \ref{fig:main_figure}, and the results are presented in Table \ref{tab:results}. 

From the H$\alpha$ line, the E+W galaxies of B14-65666 are at a redshift of 7.1513\,$\pm$\,0.0007. Galaxy E is redshifted to 7.1529\,$\pm$\,0.0004; however, galaxy W is blueshifted to 7.1481\,$\pm$\,0.0007. This implies that galaxy E and W present a velocity offset between the two components of 175\,$\pm$\,28 km\,s$^{-1}$, compatible with that observed from [O\,III]\,88\,$\mu$m and [C\,II]\,158\,$\mu$m, 177\,$\pm$\,16 km\,s$^{-1}$ (\citealt{hashimoto_2019_big}).

The H$\alpha$ line is spectrally resolved with the MIRI-MRS. It  presents an intrinsic FWHM of 407\,$\pm$\,86 km\,s$^{-1}$ for galaxies E+W and 243\,$\pm$\,41\,km\,s$^{-1}$ and 312\,$\pm$\,44\,km\,s$^{-1}$, respectively, for E and W, after deconvolving with the line spread function of the MRS (\citealt{Labiano+21}). Due to the offset in velocities the component of the E+W galaxies presents a higher FWHM than that of galaxies E and W. The observed H$\alpha$ fluxes are for E+W, E, and W:  (23.1\,$\pm$\,3.1)\,$\times\,10^{-18}$, (11.9\,$\pm$\,1.3)\,$\times\,10^{-18}$, and (8.6\,$\pm$\,1.2)\,$\times\,10^{-18}$\,erg s$^{-1}$\,cm$^{-2}$, respectively.
All results are presented in Table \ref{tab:results}. The H$\alpha$ flux for galaxies E+W is in agreement within the uncertainties of the sum of the individual galaxy fluxes. This suggests that the  H$\alpha$ emission is mainly dominated by the two galaxies excluding any relevant diffuse component.

\subsection{MIRIM and NIRCam photometry}\label{subSect:mirim_nircam_photom}

MIRIM F560W photometry for galaxies E+W and for galaxies E and W is  derived with the same apertures as for the H$\alpha$ spectra (see Fig.  \ref{fig:all} and Sect. \ref{subsect:MIRI_NIRCAM_IMAGES_MAP}). 
The local background level for each aperture is estimated in an elliptical annulus centered at the same position as the E+W galaxies' elliptical aperture and with an inner radius of 0.95\," and outer radius of 1.95\,". Additionally, we used 300 apertures on blank regions of the MIRI image with the same size as the galaxy apertures to estimate the flux uncertainties. 
We also performed the NIRCam photometry using the same aperture as for MIRIM F560W. To do so, we utilized the  NIRCam images (F115W, F150W, F200W, F277W, F356W, F444W) that are homogenized to the MIRI F560W PSF.
The final observed MIRIM F560W and NIRCam fluxes are presented in Table \ref{tab:results}. These NIRCam fluxes are lower by a factor of $\approx0.13$ than the ones obtained from previous results (\citealt{sugahara_2024_rioja}) because of the PSF homogenization from NIRCam F444W to MIRI F560W (see Sect. \ref{subsect:nircam_2}) and the different aperture sizes.

Galaxy E is an unresolved source in the NIRCam F150W image with an estimated physical size of 63 pc (see Sect. 
\ref{subSect:reff} for details). The selected aperture mainly contains emission from galaxy E; thus, we implemented aperture correction in the NIRCam and MIRIM photometry using the MIRIM F560W PSF. 
This aperture correction has a value of 1/0.67. On the contrary, galaxy W is fully resolved in the NIRCam F150W, and for that reason we did not implement aperture correction to its photometry.

\subsection{Stellar structure and effective radius}\label{subSect:reff}
We study the stellar structure of the two galaxies, E and W, individually. To achieve this, we utilized the F150W image, as it offers the best combination of signal-to-noise ratio (S/N) and spatial resolution. It presents the largest exposure time, 1675\,s, arriving at a sensitivity of 7.7\,nJy at 5\,$\sigma$ depth, which is a factor of 2\,--\,3 better than any other NIRCam band. Also, the PSF FWHM is 0.067\,", allowing us to resolve structure up to 0.35\,kpc at redshift 7.15 (\citealt{sugahara_2024_rioja}). The F150W image traces the UV emission, which is dominated by the young and attenuated stellar population. We worked with native-resolution images (before PSF-matching and resampling), and the analysis was done using the GALFIT code (\citealt{Peng+02}).

\begin{figure*} 
    \centering   \includegraphics[width=0.96\linewidth]{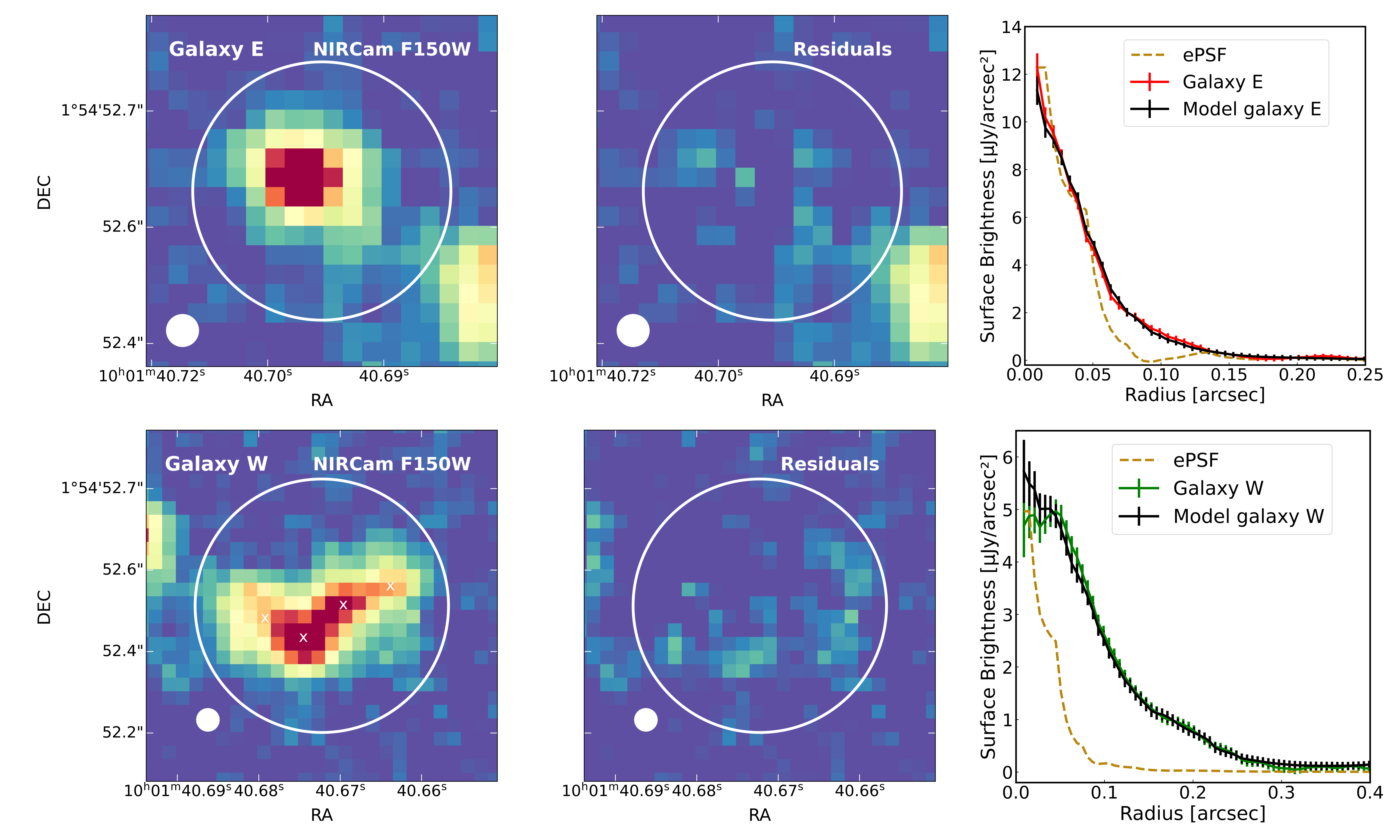}
    \caption{NIRCam F150W images and encircled light profiles for B14-65666. \textit{Top panels}: Galaxy E and residuals of the GALFIT model (modeled as one point source and one extended source). The white circular aperture represents 0.2 arcsec. \textit{Bottom panels}: Galaxy W and residuals of the GALFIT model (modeled with three point sources and one extended one, marked with white crosses). The white circular aperture represents 0.28 arcsec. The white circles shown in the bottom left of the top and bottom images represent the NIRCam F150W PSF FWHM. \textit{Top and bottom right panels}: Brown dashed curve shows the decline of the ePSF (from our observations) profile. The red and green solid lines represent the surface brightness profile of the individual galaxies E and W, respectively. The black line represents the surface brightness profile of the model of galaxies E and W, respectively, with GALFIT.}
    \label{fig:GALFIT_profiles}
\end{figure*}

Galaxy E consists of a compact core, which appears unresolved, and a nondominant extended component. To obtain the effective radius of galaxy E, we modeled the F150W data in GALFIT with a Sérsic profile (with n=0.5, which gives a Gaussian profile) to fit the diffuse emission and a point-like component (using our ePSF) to fit the more compact part of the galaxy (Fig. \ref{fig:GALFIT_profiles}). The surface brightness of galaxy E is compatible with a nuclear, point-like source plus a diffuse emission component at radii larger than 0.05\,". However, as the core remains unresolved in the F150W image, we extend our analysis to the F115W image, in which the S/N is lower, but the spatial resolution is higher than in the F150W image. The core of galaxy E also remains unresolved in the F115W image; thus, we obtained an upper limit for the effective radius of 63\,pc (0.012\,").

Galaxy W has a clumpy elongated morphology and is resolved in the F150W filter. It is well modeled by four different clumps. Three of them are unresolved point sources, which we modeled in GALFIT with point-like components. 
The other one is a more diffuse clump, associated with the brightest emission, modeled with point-like and Sérsic-profile (with an index n\,=\,0.5) components (Fig. \ref{fig:GALFIT_profiles}). When considering galaxy W as a single entity and circularizing the profile, the intrinsic effective radius is 348\,pc (0.057\,") after deconvolving with the NIRCam F150W PSF.

The merger nature of B14-65666 is supported by the morphological analysis using the \texttt{statmorph} code (\citealt{rodriguez_gomez19}). We analyzed galaxies E and W separately and find that in both cases the shape asymmetry, A$_S$ -- a reliable tracer of merging galaxies (\citealt{rinaldi_24_LRD}) -- is greater than 0.2, reinforcing the idea that each galaxy exhibits significant asymmetry. Specifically, we find A$_S$ of 0.35 and 0.80 for galaxies E and W, respectively. We also tested the multimode-intensity-deviation (MID) statistics, which are useful for detecting multicomponent systems, such as the double-nucleus of a late-stage merger, highly disordered post-merger remnants, galaxies with bright star-forming clumps in rest-UV emission, or an apparently single galaxy with an extended emission component(s) (\citealt{freeman_2013_MID}). 
M (multimode statistics) quantifies how a galaxy’s light is distributed into distinct regions, and D (deviation statistic) measures the offset between the brightest region and the overall centroid. M values approaching 1 are likely to indicate a double nucleus, while values close to zero are interpreted as indicating a single source; D values close to zero indicate symmetric and ordered morphologies such as spheroids and disks.
In the case of galaxy E, M\,$\approx$\,0.0 and D\,$\approx$\,0.1. This corroborates the interpretation that galaxy E is a single (compact spheroidal) source. 
In the case of galaxy W, M\,$\approx$\,0.4 and D $\approx$\,0.3, indicating that galaxy W is composed of at least two clumps, with one of them dominating the emission.

\section{Results and discussion}\label{Sec:results_dis} 

We detected and spatially resolved the rest-frame optical continuum and H$\alpha$ emission in the system B14-65666 at $z$\,=\,7.15 with MIRI imaging and spectroscopy. The new MIRI data are combined with ancillary NIRCam imaging and ALMA spectroscopy to obtain the physical properties of galaxies E and W, and of the entire system: star formation rate, instantaneous SFR, ionizing photon-production efficiency, H$\alpha$ equivalent width, dynamical mass, stellar-mass surface density, and SFR surface density.

\begin{figure*} 
    \centering   \includegraphics[width=0.96\linewidth]{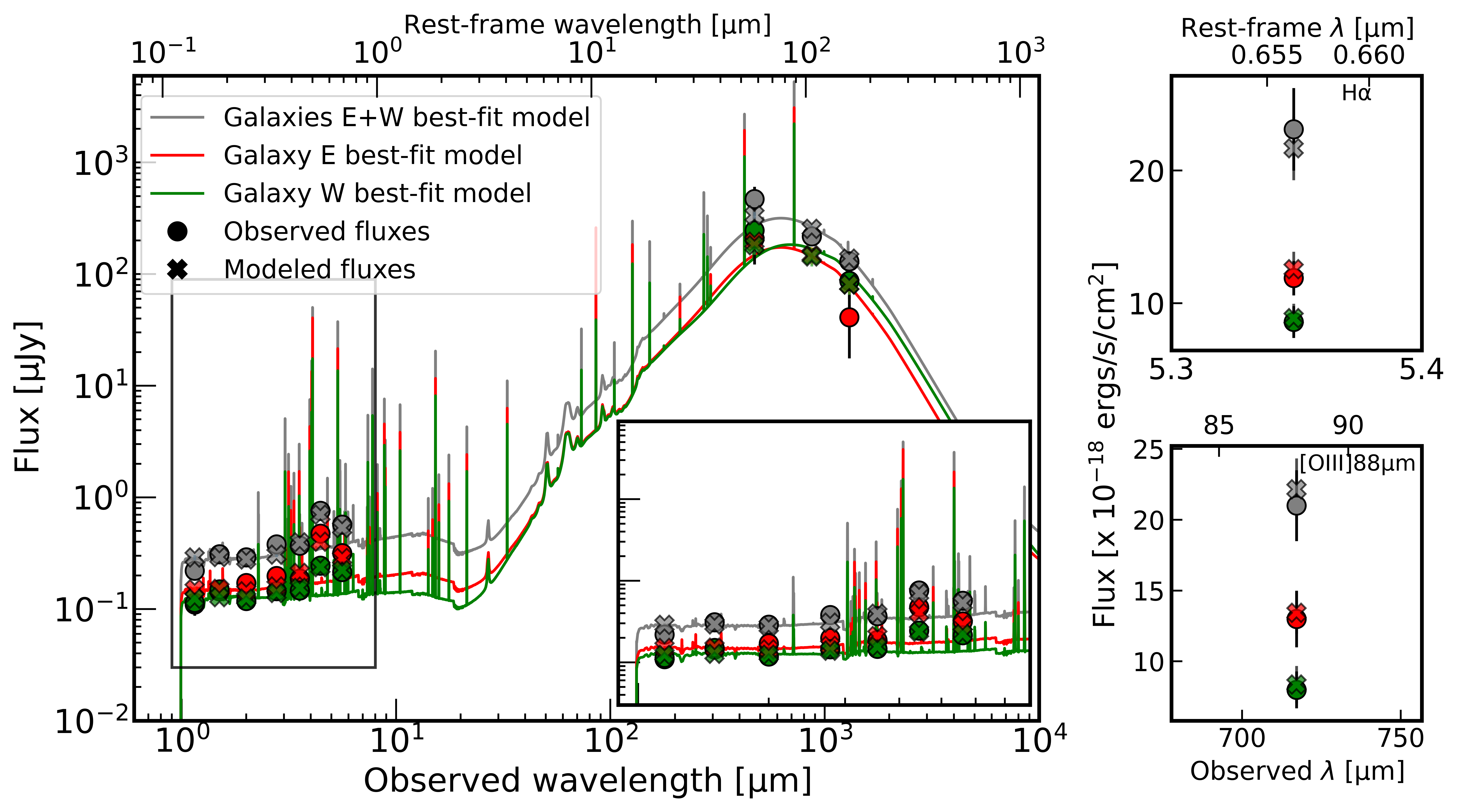}
    \caption{Observed near-infrared-to-millimeter SEDs and best-fit models derived from CIGALE SED fitting analysis of galaxies E+W, E, and W. Gray line: Best-fit model for the galaxies E+W. Red line: Best-fit model for galaxy E. Green line: Best-fit model for galaxy W. The circles represent the observed photometric fluxes for galaxies E+W, E, and W, with the associated uncertainties given in Table \ref{tab:results}. The crosses represent the modeled fluxes. Also indicated are the H$\alpha$ observed and predicted fluxes (top right) and the corresponding [O\,III]\,88\,$\mu$m fluxes (bottom right).} 
    \label{fig:sed_cigale}
\end{figure*}

\begin{table*}[]
\centering
\caption{Physical parameters for B14-65666 derived from SED fitting analysis with CIGALE and with observational data.}
\label{tab:physical_parameters}
\begin{tabular}{lcccc}
\hline
Parameter           & Galaxies E+W      & Galaxy E               & Galaxy W                                                            \\ \hline

M$_{\star}$ [$\times 10^{8}$ M$_{\odot}$$_{total}$ ]$^{(a)}$         & $28 \pm 16   $      & $15 \pm9  $              & $ 8 \pm4$\\

M$_{dyn}$ [$\times 10^{9}$ M$_{\odot}$$_{total}$ ]$^{*}$         & -     & $1.5 \pm0.3  $              & $ 8.5 \pm2.4$ \\

Age$_{young}$  [Myr]$^{(a)}$ &$ 7\pm 3 $      & $6 \pm3 $          &$ 8\pm3    $                                                    \\
Age$_{main}$      [Myr]$^{(a)}$      & $214\pm150$ & $224\pm150$ & $158\pm155$
 \\

 Metallicity  [Z$_{\odot}$]$^{(a)}$                  &      $0.25\pm  0.11$          &          $0.24\pm 0.12$             &       $0.31\pm 0.15$                  \\ 

EW$_{0}$(H$\alpha$) [$\AA$]$^{*}$ & - & $832 \pm 100$ & $536 \pm 78$\\

EW$_{0}$(H$\alpha$) [$\AA$]$^{(a)}$              & $800 \pm 159$           & $880 \pm180$               & $833 \pm180 $                                                           \\
                   
 $L_{dust}$ [$\times 10^{11}$\,L$_{\odot}$]$^{(a)}$                   &     $4.8\pm0.4 $              &                     $2.6\pm0.5 $   &    $2.5\pm0.5 $ 

 \\
 $L_{UV}$ [$\times 10^{21}$ W$Hz^{-1}$]$^{(a)}$                   &     $22.6\pm1.1$              &                     $11.8\pm0.7$   &    $9.7\pm0.6$ 
  \\
 $A_{FUV}$[mag]$^{(a)}$                   &     $1.4\pm0.1$              &                     $1.5\pm0.2$   &    $1.6\pm0.2$ 
  \\
 $A_{V}$ [mag]                  &  - &                     $1.5\pm0.4$   &    $0.1 ^{+0.5}_{-0.1}$ 
 
 \\
 $\log(\zeta_\mathrm{ion}$ [Hz erg$^{-1}$])$^{*}$ & - & $ 25.1\pm0.1 $ & $ 25.5\pm0.1 $ \\
SFR    [M$_{\odot} $yr$^{-1}$]$^{(a)}$    &    $127\pm66 $  & $84\pm49 $   &  $50\pm12 $ \\
SFR$_\mathrm{H_{\alpha},Z=Z_\odot}$\,[M$_{\odot}$\,yr$^{-1}$]$^{*}$       & - & $ 76 \pm 8$ & $30 \pm 4$ \\
SFR $_\mathrm{H_{\alpha},Z=0.29Z_\odot}$\,[M$_{\odot}$\,yr$^{-1}$]$^{*}$       & -& $ 46 \pm 5$ & $18 \pm 3$ \\
SFR$_\mathrm{[C\,II]158 {\mu m}}$\,[M$_{\odot}$\,yr$^{-1}$] &   140 $\pm$ 130   &      68 $\pm$ 64   &    53 $\pm$ 50    &  \\
SFR$_\mathrm{H_{\alpha}+TIR}$\,[M$_{\odot}$\,yr$^{-1}$]$^{*}$                                                 &     107 $^{+35}_{-16}$  &    50 $^{+14}_{-9}$    &      54 $^{+27}_{-17}$  &  \\
SFR$_\mathrm{UV}$\,[M$_{\odot}$\,yr$^{-1}$]$^{*}$                                         &  72 $\pm$ 4  &   
41 $\pm$ 3    & 37 $\pm$ 2 \\
SFR$_\mathrm{UV+TIR}$\,[M$_{\odot}$\,yr$^{-1}$]$^{*}$                                                      &  147$^{+62}_{-22}$  &    66  $^{+24}_{-15}$   &    88  $^{+43}_{-19}$   &  \\

\hline
\end{tabular}
\tablefoot{($^{a}$) Obtained with SED fitting from CIGALE. ($^{*}$)The derived parameters are corrected for dust attenuation (see Sect. \ref{subSec:attenuation}).}
\end{table*}

\subsection{Spectral energy distribution and star formation history}\label{subSect:SED} 

We performed an SED fitting using \texttt{CIGALE} (\citealt{noll_2009_analysis}; \citealt{boquien_2019_cigale}) for galaxies E+W as well as for the two individual components. The SED covers the 1\,$\mu$m to 1.3\,mm observed wavelength range, combining JWST photometric data (NIRCam and MIRIM) with ALMA continuum imaging (\citealt{hashimoto_2019_big}; \citealt{sugahara_2021_big}), along with our H$\alpha$ and [O\,III]\,88\,$\mu$m (\citealt{hashimoto_2019_big}) line fluxes. The redshift used in the analysis is the one derived spectroscopically from the H$\alpha$ line.

We made different assumptions to represent the star formation history of the system. The best results are determined from \texttt{CIGALE} fitting by minimizing the $\chi$$^{2}$ value and ensuring the best agreement with observed photometry and line fluxes. These are obtained when considering two components: an instantaneous burst with ages ranging from 30 to 500 Myr to model the mature stellar population; and a constant burst with ages lower than 10 Myr to model the young stellar population.
We used the \cite{BruzualandCharlot+03} stellar population models and the \cite{Chabrier+03} initial mass function (IMF). The nebular continuum and emission lines were modeled using electron densities from 100 to 1000 cm$^{-3}$ and the ionization parameter (logU) ranging from -3.5 to -1.5. The metallicity was treated as a free parameter, ranging from 0.03\,Z$_{\odot}$ to 0.7\,Z$_{\odot}$, as previous studies showed that it varies in galaxies E and W (\citealt{sugahara_2024_rioja}; \citealt{gareth_2024}). We used the dust attenuation law from \cite{Calzetti+00}, and the dust emission was modeled using the \cite{Draine2014} models.

The best-fit models for galaxies E+W and for galaxies E and W separately are shown in Fig. \ref{fig:sed_cigale}. These models reveal that B14-65666 is dominated by a young stellar burst characterized with a total SFR of 127\,$\pm$\,66\,M$_{\odot}\,$yr$^{-1}$ and a mass for the young stellar component of (8\,$\pm$\,1)\,$\times$\,10$^{8}$
M$_{\odot}$. Galaxies E and W have SFRs of 84\,$\pm$\,49 and 50\,$\pm$\,12\,M$_{\odot}\,$yr$^{-1}$, respectively, and show the same young stellar component with a stellar mass of (4\,$\pm$\,1)\,$\times$\,10$^{8}$\,M$_{\odot}$. The SED fitting results indicate the presence of a mature stellar population in B14-65666 with a stellar mass of around 0.5\,$-$\,2\,$\times$\,10$^{9}$\,M$_{\odot}$ and average ages about 200 Myr. SED fitting results are presented in Table \ref{tab:physical_parameters}.
We conclude from the SED fitting that the B14-65666 system consists of two galaxies experiencing starbursts characterized by a similar age 
(see Sect. \ref{subSect:simulations} for a more detailed discussion). Also, despite the compactness of galaxy E (see Sect. \ref{subSect:mass-size}), the SED based on the NIRCam and MIRI imaging and covering the rest-frame 0.16\,--\,0.78\,$\mu$m
does not correspond to that of the average little red dots at redshifts $\sim$\,6\,--\,8 (\citealt{perez_2024}; \citealt{kokorev_24}; \citealt{rinaldi_24_LRD}).

\subsection{Nebular and stellar attenuation} \label{subSec:attenuation}

Dust attenuation plays a key role in shaping the observed properties of B14-65666, as revealed by MIRI, NIRCam, and ALMA observations. Combining the MRS H$\alpha$ and the NIRSpec H$\beta$ fluxes\footnote{We implemented aperture corrections (AC\,=\,1/0.79) using the NIRSpec IFS PSF to match the fluxes presented in \cite{gareth_2024} to the larger aperture of our MRS measurements.}, the H$\alpha$/H$\beta$ ratio gives observed values of 4.5$\,\pm$\,0.6 and 2.9\,$\pm$\,0.4 for galaxies E and W, respectively. We note that these values correspond to the narrow component of the line emission for both galaxies as the broad component present on the H$\beta$ line (\citealt{gareth_2024}) is not detected in H$\alpha$ with the present data (see Sect. \ref{subSect:kinematics}).
Assuming a case B recombination, T$_{e}$\,$\approx$\,1.5\,$\times$\,$10^{4}$\,K and n$_{e}$\,=\,$10^{3}$\,cm$^{-3}$, the theoretical ratio is 2.8. Using \cite{Cardelli+89}, we obtain that the nebular visual extinction ($A_{V}$) is $A_{V}$\,=\,1.5\,$\pm$\,0.4 mag for galaxy E, 
and it is $A_{V}$\,=\,$0.1 ^{+0.5}_{-0.1}$ mag for galaxy W. 
Thus the nebular attenuation is higher for galaxy E, in agreement, within the uncertainties, with the values of $A_{V}$\,=\,1.6\,$\pm$\,0.3 mag for galaxy E and $A_{V}$\,=\,0.7\,$^{+0.3}_{-0.2}$ mag for W from \cite{sugahara_2024_rioja} (obtained with SED fitting and applying the conversion factor to nebular attenuation, f\,=\,0.44; \citealt{calzetti97}).

We also calculated the dust attenuation in the FUV, $A_{FUV}$, with the IR-to-UV luminosity ratio (IRX\,=\,$L_{IR}$/$L_{UV}$), as this is a robust tracer of $A_{FUV}$. 
Taking the value of $log_{IRX}$\,=\,0.7 for the total galaxy (\citealt{sugahara_2024_rioja}), we adopted expression (5) from \cite{meurer99} assuming the bolometric correction for the total light emitted by dust is 1, and for the total light emitted by stars, UV, is 1.68 (\citealt{meurer99}). We obtain a value of $A_{FUV}$\,=\,1.50, which is compatible with that obtained with our SED fitting for galaxies E+W, $A_{FUV}$\,=\,1.41\,$\pm\,0.09$ mag (\citealt{Calzetti+00}).

\subsection{Star formation rate and burstiness}\label{subSec:SFR}

We derived the recent (i.e., less than 10 Myr) star formation rate directly from our H$\alpha$ measurements, assuming solar (to facilitate comparison with other star formation tracers at other wavelengths) and subsolar metallicity (0.29 Z$_{\odot}$, \citealt{gareth_2024} and see Table \ref{tab:physical_parameters}). We applied the corresponding SFR-to-H$\alpha$ conversion factor for each metallicity, which is approximately 1.7 times lower in the subsolar case (\citealt{Reddy+18}). The final SFR values are corrected for attenuation using the H$\alpha$/H$\beta$ ratios (see previous section). We obtain SFRs of $76\,\pm\,8$ ($46\,\pm\,5$) M$_{\odot}$\,yr$^{-1}$ and $30\,\pm\,4$ ($18\,\pm\,3$) M$_{\odot}$\,yr$^{-1}$ for galaxies E and W and solar (subsolar) metallicities, respectively.

Additional estimates of the SFR can be obtained directly from far-infrared emission lines or using a combination of multiwavelength tracers. The [C\,II]\,158\,$\mu m$ emission line is a tracer of SFR in low-redshift, star-forming, and composite SF+AGN galaxies (\citealt{DeLooze+14}); it is confirmed for high-redshift galaxies (\citealt{Schaerer+20}; \citealt{carninai_ferrara_2020}). Adopting the ancillary [C\,II]\,158\,$\mu$m luminosities derived from ALMA (\citealt{hashimoto_2019_big}), values similar to those derived from our H$\alpha$ measurements are obtained, within the uncertainties (see Table \ref{tab:physical_parameters}). The large uncertainty in the value of the SFR derived from the [C\,II]\,158\,$\mu$m emission line is due to the nature of this emission, which could include extended or diffuse regions not directly linked to active star formation. Also, to trace less massive stars we used the SFR(UV) for solar metallicities and considering a constant SFH over 100 Myr (\citealt{Calzetti+13}).

Other SFR estimates involve the use of hybrid indicators such as the UV+TIR (total infrared) luminosity (\citealt{Kennicutt-Evans+12}), or H$\alpha$+TIR (\citealt{catalntorrecilla_2015_star}; \citealt{Calzetti+13}), that capture both obscured and unobscured star formation over longer periods of time (up to 100 Myr). Assuming a simple energy balance  (\citealt{kennicutt_2009_dustcorrected}), and considering the far-UV luminosity derived from our SED fitting ($L(\text{UV})$\,=\,(1.15\,$\pm$\,0.11)\,$\times\,10^{11}$\,$L_\odot$), and the TIR luminosity $L(TIR)$\,=\,(1.05\,$\pm$\,0.2)\,$\times\,10^{12}$\,$L_\odot$ (\citealt{hashimoto_2019_big}), we obtain an SFR consistent with our extinction-corrected H$\alpha$ estimates. The SFR derived from the H$\alpha$ and TIR luminosities are also in good agreement with those obtained from UV+TIR (see Table \ref{tab:physical_parameters}). The SFRs derived in this analysis (70\,--\,140\,M$_{\odot}$\,yr$^{-1}$) are lower than the SED-based SFR reported by \cite{sugahara_2024_rioja}, SFR\,=\,225\,$^{+71}_{-56}$\,M$_{\odot}$\,yr$^{-1}$, and the (narrow+broad) H$\beta$-based SFR from \cite{gareth_2024}, SFR$_{H\beta}$\,=\,308\,$\pm$\,39\,M$_{\odot}$\,yr$^{-1}$. This difference is attributed to variations in aperture sizes where the contribution of the extended emission (H$\beta$) is significant, as well as other factors such as different estimates of the internal dust attenuation. Additionally, in the case of the H$\beta$-based SFR, the presence of both narrow and broad components in the H$\beta$ line emission also contributes to the higher SFR value by a factor of 2 (\citealt{gareth_2024}).

A useful diagnostic for assessing the recent star formation in high-${\it z}$ galaxies is the ratio of the SFR(H$\alpha$)-to-SFR(UV), which is referred to as the burstiness parameter and probes star formation over the last 10\,Myr relatively to the last 100\,Myr. If the ratios are above 1, it indicates that the galaxy is in a bursty phase (\citealt{Atek+23}). We obtain a value of the ratio SFR(H$\alpha$)-to-SFR(UV) for solar metallicities of 1.8\,$\pm$\,0.1 and 0.8\,$\pm$\,0.1 for galaxies E and W, respectively (we would expect similar values for subsolar metallicities). When compared to the range of burstiness ratios reported by \citet{Atek+23} (0.7\,<\,\textit{z}\,<\,1.5), values above or close to unity are typical in low-mass galaxies (M$_{\star}$<$10^{9}$\,M$_{\odot}$) and below unity in high-mass galaxies (M$_{\star}$>$10^{9}$\,M$_{\odot}$). Thus, our values suggest that galaxy E shows signs of bursty activity, forming stars at nearly twice the average rate of the past 100 Myr; it is  in a mild bursty phase,  while galaxy W forms stars at a rate closer to that if this past 100 Myr; i.e., experiencing a more continuous burst. This aligns with the results from \cite{navarro-carrera24}, whose authors analyzed $\sim4500$ H$\alpha$ emitters across $\textit{z}$\,$\sim$\,2\,--\,6.5 using JADES data, observing a correlation between enhanced SFR(H$\alpha$)/SFR(UV) and EW$_0$(H$\alpha$), reinforcing the scenario where galaxies with above-unity SFR(H$\alpha$)/SFR(UV) experience a bursty phase in the last $\sim 50$\,Myr.

\begin{figure} 
    \centering   \includegraphics[width=0.96\linewidth]{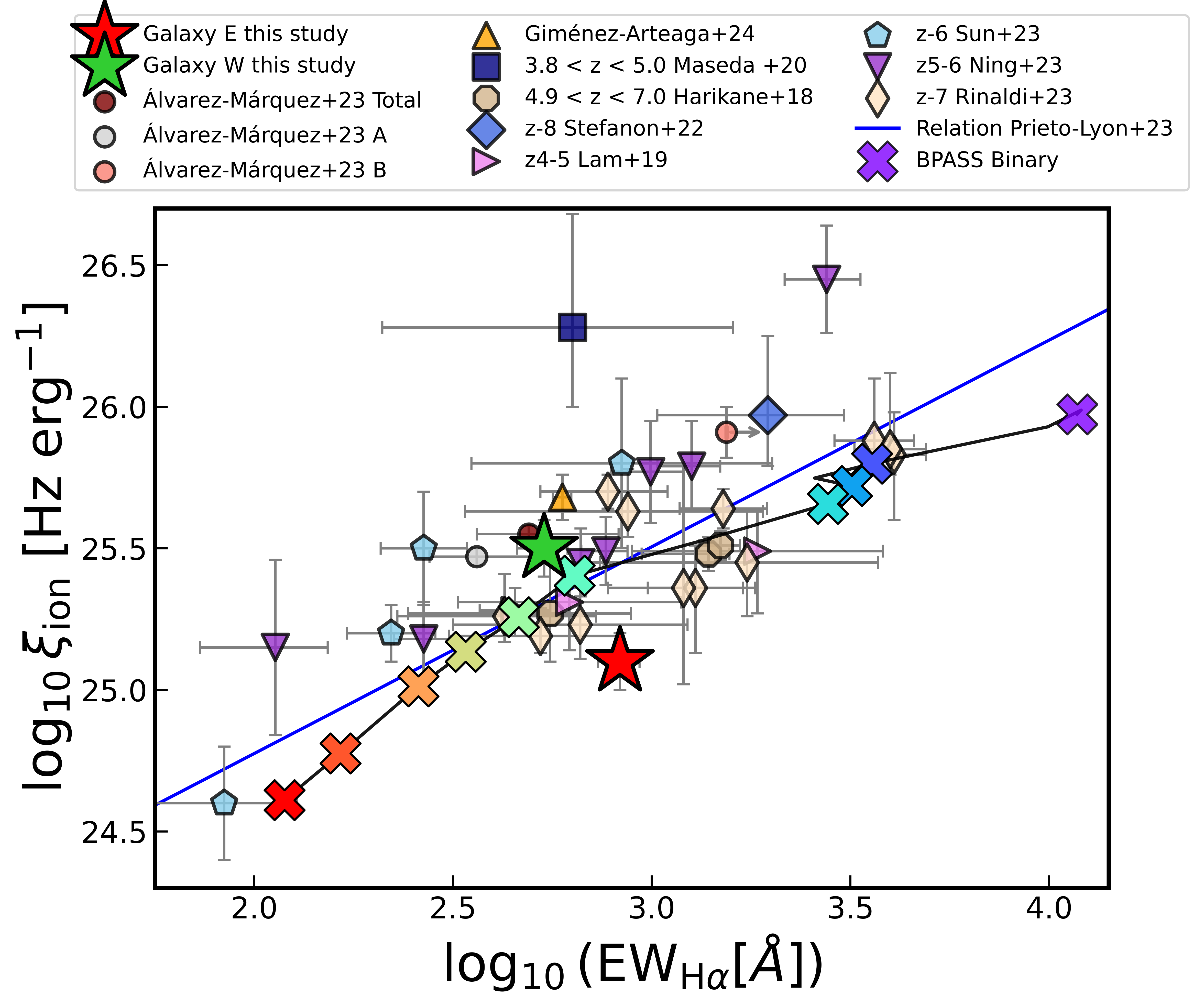}
    \caption{$\log(\zeta_\mathrm{ion})$ as function of H$\alpha$ equivalent width, both of them corrected from dust attenuation. The red star represents galaxy E of B14-65666 and the green star galaxy W of B14-65666. The blue line represents the relation for galaxies at redshifts 3 to 7 (\citealt{Prieto-Lyon+23}). B14-65666 is compared with the galaxy MACS1149-JD1 (\citealt{alvarez-marquez+23c}), RXCJ0600-z6-3 (\citealt{gimenez-arteaga_2024}), mean of 102 Lyman-break galaxies at redshift 8 (\citealt{Stefanon+22}), MIDIS H$\alpha$-emitters at redshifts 7-8 (\citealt{Rinaldi+23}), faint Ly$\alpha$ emitters (LAEs) at redshifts 4-5 (\citealt{Lam+19}), LAEs at 3.8-5 (\citealt{Maseda+20}), a LAE subsample at redshift 4.9 (\citealt{harikane2018}), line emitters at redshifts 6.11\,--\,6.35 (\citealt{sun_2023}), and LAEs at redshift 6 (\citealt{Ning+23}). We also present the BPASS models of a binary stellar population for an instantaneous burst with ages from 1 to  20 Myr 
   (crosses; from red to blue, the ages are 20, 16, 13, 10, 8, 6, 4, 3, 2, and 1 Myr and are connected with the solid black line) 
  (\citealt{Stanway+Eldridge-23}).}
    
    \label{fig:log_Ion-EW}
\end{figure}

\subsection{Equivalent width of H$\alpha$ and ionizing photon-production efficiency}\label{subSec:ew}

To derive the rest-frame equivalent width of H$\alpha$ (EW$_0$(H$\alpha$)) for galaxies E+W and galaxies E and W, we make use of the F560W and H$\alpha$ fluxes measured in our MIRIM and the MRS observations, respectively. First, we obtain rest-frame optical continuum emission by subtracting the contribution of the H$\alpha$ line (Sect. \ref{subSect:halpha_spectra}) to the F560W flux (Sect. \ref{subSect:mirim_nircam_photom}). The EW$_0$(H$\alpha$) is then derived as the ratio of the H$\alpha$ to the continuum flux under the H$\alpha$ line. We obtain rest-frame equivalent widths of 832\,$\pm$\,100\,\r{A} and 536\,$\pm$\,78 \r{A} for galaxies E and W, respectively.  
The difference in the EW$_0$(H$\alpha$) of the two galaxies is attributed to a different star formation history in the two galaxies with ionizing young stars and more mature, non-ionizing stellar populations contributing in different proportions to the continuum emission. 
According to our SED analysis, galaxy E, with a larger EW$_0$(H$\alpha$) value, has a slightly younger stellar population (6 Myr) relative to galaxy W (8 Myr).

The ionizing photon production efficiency, $\log(\zeta_\mathrm{ion}$), is the ratio of the ionizing to the non-ionizing UV flux. Assuming recombination, $\log(\zeta_\mathrm{ion}$) is empirically obtained as the ratio of the number of ionizing photons ($N_{LyC}$) derived from the H$\alpha$ line (after extinction correction) and the luminosity at 1500 \r{A}. Assuming zero escape fraction, a dust attenuation as derived in Sect. \ref{subSec:attenuation}, and a temperature of 1.5\,$\times$ 10${^4}$\,K (\citealt{alvarez-marquez+23c}) the integrated H$\alpha$ luminosity provides a total number of ionizing photons ($N_{LyC}$) equal to $1.1\times10^{55}$\,ph\,s$^{-1}$ and $4.1\times10^{54}$\,ph\,s$^{-1}$ for galaxies E and W, respectively. Different approaches to dust correction can yield significantly different values of $\zeta_\mathrm{ion}$ introducing uncertainty in its determination and affecting the interpretation of its physical implications (\citealt{matthee17}).
Thus, correcting the UV and H$\alpha$ luminosities, we obtain ionizing photon-production efficiencies $\log(\zeta_\mathrm{ion}$) equal to $25.1\pm0.1$ and $25.5\pm0.1$ Hz erg$^{-1}$ for galaxies E and W, respectively. This difference in the $\zeta_\mathrm{ion}$ between the two galaxies of the system can be interpreted as partly due to the age difference in the young stellar populations in the two galaxies. Models indicate that in addition to the stellar continuum, there is a strong nebular contribution due to the two-photon continuum at 1500$\AA$. This contribution can be close to 50\% of the total for very young populations (1 Myr), decreasing with age and becoming almost negligible for 10-Myr-old populations (\citealt{katz_2024}). Thus, the impact of this nebular continuum would be to increase the 1500\,$\AA$ luminosity and consequently decrease the corresponding $\log(\zeta_\mathrm{ion})$ by factors of 0.15 to 0.04 for ages of 1 to 8 Myr. In addition, residual differential extinction in the two galaxies could also play a role due to uncertainties in the derivation of the extinction in the UV.

We present the previous results in the $\log(\zeta_\mathrm{ion})$\,--\,EW$_0$(H$\alpha$) plane (see Fig. \ref{fig:log_Ion-EW}). Additional samples of galaxies at redshifts 4 to 9 are also presented for comparison. These include Lyman-alpha emitters (LAE) at redshifts 3.8-6 (\citealt{Maseda+20}; \citealt{Ning+23}; \citealt{Lam+19}; \citealt{harikane2018}), H$\alpha$ and [O\,III] emitters at redshifts 6.1 to 8 (\citealt{sun_2023}; \citealt{Rinaldi+23}), and the mean of Lyman-break galaxies at redshift 8 (\citealt{Stefanon+22}). Individual galaxies with resolved integral-field spectroscopy are also shown (RXCJ0600-2007 at $z$\,=\,6.07, \citealt{gimenez-arteaga_2024} and MACS1149-JD1 at $z$\,=\,9.1, \citealt{alvarez-marquez+23c}). While galaxies E and W forms stars at a rate much higher than the average galaxy at redshift 7, they have $\log(\zeta_\mathrm{ion})$ and EW$_0$(H$\alpha$) in the range of high-${\it z}$ galaxies. These values are consistent with the predicted values for young stellar populations. The Binary Population and Spectral Synthesis (BPASS; \citealt{Eldridge+Stanway+20}) model demonstrates that the $\log(\zeta_\mathrm{ion}$) and the EW$_0$(H$\alpha$) of a single-aged, instantaneous-starburst binary population increases as the age of the population decreases (see Fig. \ref{fig:log_Ion-EW}), with values compatible with those observed for young populations with ages from 1 to 20 Myr. The production of ionizing photons also depends on the presence of massive stars and therefore the IMF (\citealt{stanway_16}). Values for the logarithm of the photon production efficiency above 25.5 Hz\,erg$^{-1}$ can only be obtained with massive bursts, $10^{7}$\,M$_{\odot}$ due to the stochasticity of the formation of massive stars (\citealt{Stanway+Eldridge-23}). The galaxies in the B14-65666 system and the high-{\it z} galaxies identified with the JWST should have higher young stellar masses.

\begin{figure*} 
    \centering   \includegraphics[width=0.96\linewidth]{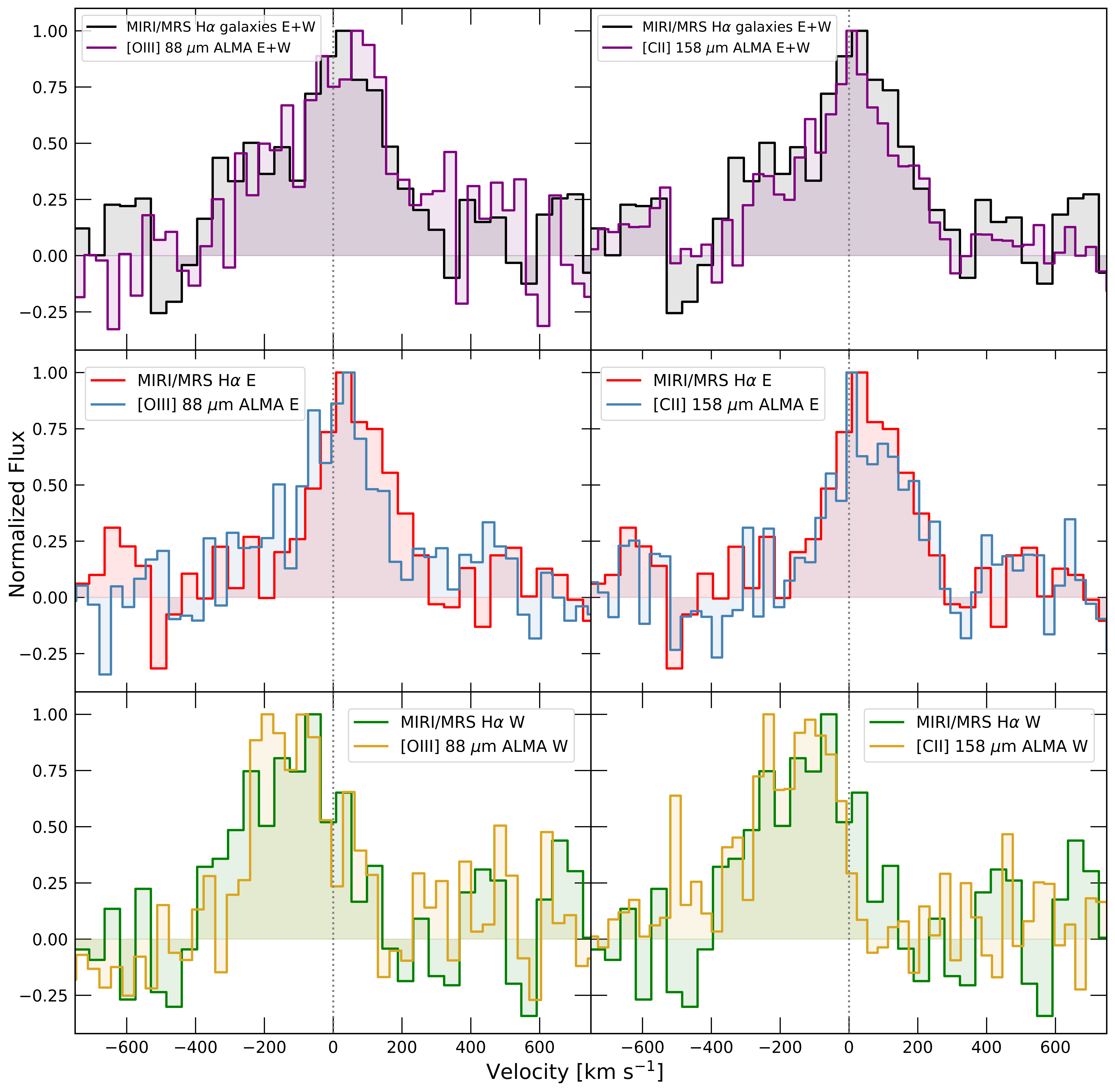}
    \caption{Comparison of integrated H$\alpha$ emission-line profile, with the ALMA spectrum of the [O\,III]\,88\,$\mu$m line and [C\,II]\,158\,$\mu$m line (\citealt{hashimoto_2019_big}) for galaxies E+W, for galaxy E, and for galaxy W, respectively, from top to bottom. The systemic velocity corresponds to redshift 7.1513.}
    \label{fig:oiii_cii_A_B}
\end{figure*}

\begin{figure} 
    \centering   \includegraphics[width=0.9\linewidth]{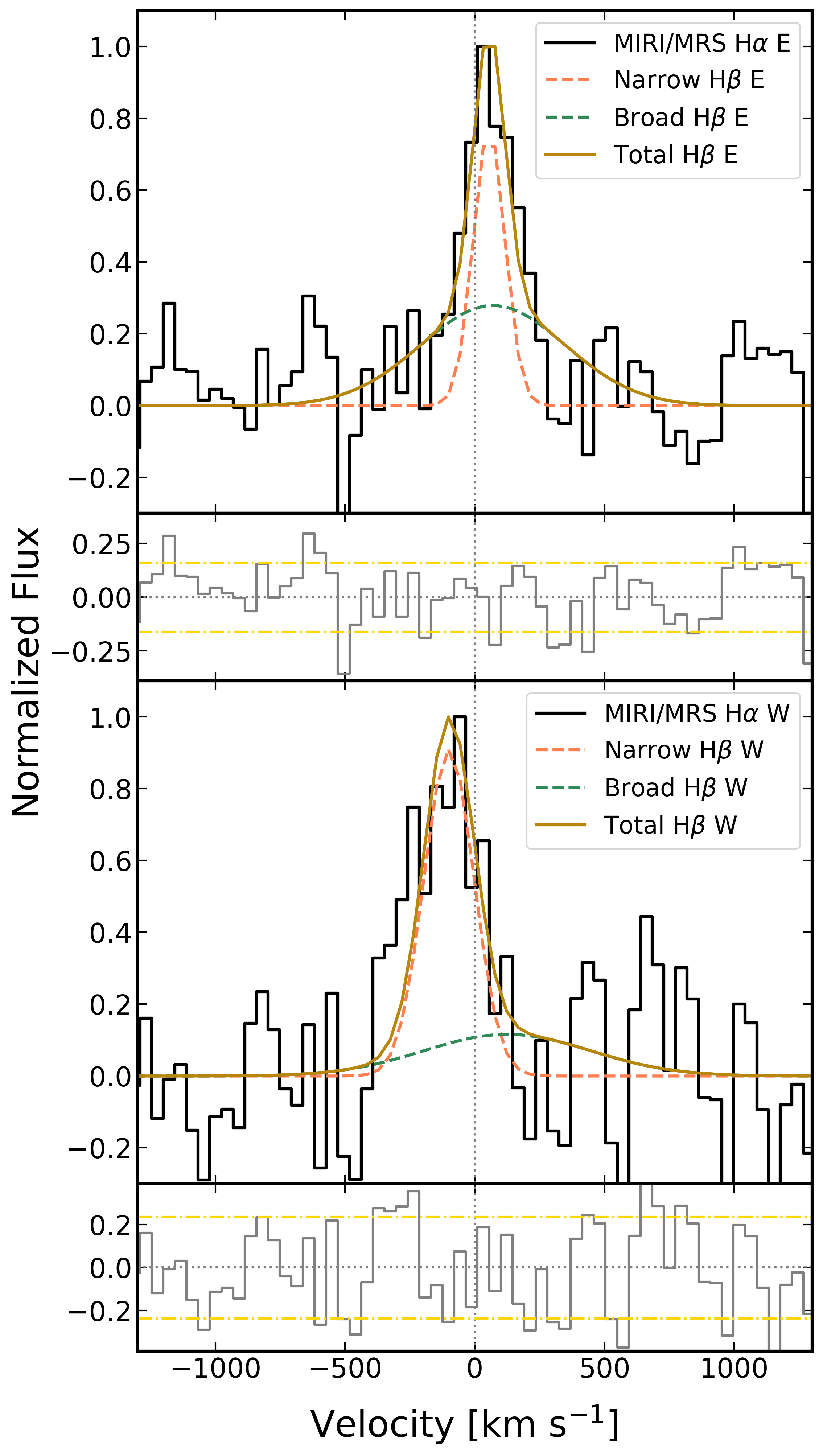}
    \caption{Direct comparison of MRS-observed H$\alpha$ emission-line profile for galaxies E (top) and W (bottom), with the rescaled H$\beta$ narrow- and broad-line components identified in lower spectral resolution NIRSpec IFS (\citealt{gareth_2024}). The presence of the narrow plus broad components could be compatible with the observed H$\alpha$ line profile (see Sect. \ref{subSect:kinematics} for a detailed discussion). The systemic velocity corresponds to redshift 7.1513. The residuals are shown in the bottom plots, together with the $1\sigma$ noise level (yellow).}
    \label{fig:hbeta}
\end{figure}

\subsection{Kinematics of the ionized ISM}\label{subSect:kinematics} 

The kinematics of the ionized gas can be measured with the H$\alpha$ emission-line profile in our MRS data. Based on the integrated H$\alpha$ line, we measure a redshift of $z$\,=\,7.1513\,$\pm$\,0.0007 for the B14-65666 system, compatible with the values obtained with the FIR emission lines [O\,III]\,88\,$\mu$m and [C\,II]\,158\,$\mu$m ($z$\,=\,7.1521\,$\pm$\,0.0004, \citealt{hashimoto_2019_big}) and optical [O\,III] line ($z$\,=\,7.1517\,$\pm$\,0.0001, \citealt{gareth_2024}). 
The two galaxies of the B14-65666 system have a relative velocity of 175\,$\pm$\,28\,km\,s$^{-1}$ along our line of sight as derived from the centroid of the H$\alpha$ line emission, showing it is a galaxy merger.

The kinematics of the ionized gas traced by H$\alpha$ appears to show differences between the two galaxies. 
Specifically, we find that the FWHM for galaxy W is higher than for galaxy E, with values of 312\,$\pm$\,44\,km\,s$^{-1}$ and 243\,$\pm$\,41\,km\,s$^{-1}$, respectively. These values agree within the uncertainties (from those measured) with the far-infrared emission lines, [C\,II]\,158\,$\mu$m and [O\,III]\,88\,$\mu$m. The [C\,II]\,158\,$\mu$m emission has similar kinematics in both galaxies (FWHM of 288\,$\pm$\,41\,km\,s$^{-1}$ for galaxy W and 276\,$\pm$\,34\,km\,s$^{-1}$ \footnote{The FWHM value for clump E has been recalculated and differs from that reported in \cite{hashimoto_2019_big}, as confirmed through a private communication with the authors.} for galaxy E). For [O\,III]\,88\,$\mu$m, galaxy E shows a broader line (322\,$\pm$\,41\,km\,s$^{-1}$) than galaxy W (267\,$\pm$\,42\,km\,s$^{-1}$). In Fig. \ref{fig:oiii_cii_A_B}, we present the comparison of these three lines for galaxies E+W and E and W. The observed differences in the two galaxies indicate the complex structure of the velocity field in the different phases of the ISM. While the H$\alpha$ line traces the overall kinematics of the ionized gas, the [C\,II]\,158\,$\mu$m line is closely associated with the neutral gas and the [O\,III]\,88 $\mu$m with the high excitation gas. Simulations also show that different tracers are sensitive to different kinematic properties (\citealt{kohandel}). In addition, the continuum imaging (\citealt{sugahara_2024_rioja}) of both galaxies shows a different morphology; galaxy E as a very compact source, while galaxy W shows a more extended structure, in particular in the rest-UV light, reminiscent of tidal tails.
Moreover, low-spectral-resolution (R\,$\approx$\,1000) integral-field spectroscopy with NIRSpec suggests the presence of a narrow and a broad kinematic component in both galaxies and in all optical emission lines tracing the ionized gas (\citealt{gareth_2024}). These two components are potentially associated with the motions in the host galaxy (narrow) and the presence of high-velocity gas outflows (broad). The width of the narrow component is only 144\,$\pm$\,11\,km\,s$^{-1}$ for galaxy E and 227\,$\pm$\,6\,km\,s$^{-1}$ for galaxy W. These are significantly narrower than the values derived from our H$\alpha$ line and far-infrared emission lines, obtained with a better spectral resolution but lower S/N. We tested whether the observed H$\alpha$ line profile could be compatible  with the combination of the narrow and broad (653\,$\pm$\,10\,km\,s$^{-1}$ for galaxy E and 755\,$\pm$\,23\,km\,s$^{-1}$ for galaxy W) line components identified with NIRSpec. Our tests (see Fig. \ref{fig:hbeta}) indicate that the narrow+broad (N+B) component could be slightly compatible with our observed H$\alpha$ line profiles. However, the simulated H$\alpha$ N+B line profile appears to have a flux excess in the spectral range of the line (i.e., residuals mostly below zero) for galaxy E, while the faint broad component in galaxy W is close in velocity to the peak emission in galaxy E. Since the two galaxies are separated by only 0.44 arcsec, this secondary emission in W could be due to a contamination from the brighter emission from E (\citealt{gareth_2024}).
Further higher spectral resolution data with NIRSpec, and higher S/N MRS spectra, would be required to confirm the presence of such a broad line emission in the two galaxies.

The consistency of the trends between our results and the ancillary data provides robust evidence of the dynamic differences between galaxies E and W. The broader profiles observed in galaxy W for both H$\alpha$ and [C\,II]\,158\,$\mu$m may indicate enhanced turbulence or kinematic effects, potentially driven by interactions or feedback processes, as well as a possibility of a connection with outflows. On the other hand, the higher FWHM in [O\,III]\,88 $\mu$m for galaxy E suggests that highly ionized gas dynamics differ from those of lower ionization states, pointing to localized physical conditions or energetic phenomena. 
Also, dust attenuation could impact the interpretation, in particular in galaxy E where a large visual extinction is measured. 
The optical lines are more susceptible to dust extinction, which can affect measurements of line widths and kinematic features, whereas far-infrared lines are less impacted by such effects.

\subsection{Stellar and dynamical mass}\label{subSect:morph} 

Although the line profiles of the different emission lines suggest that the gas is not fully (dynamically) relaxed, here we obtain an upper limit of the dynamical mass of each of the galaxies assuming the system is close to relaxation and characterized by the random motions traced by the ionized gas velocity dispersion; i.e., not dominated by rotational motions. Under this working hypothesis, we followed previous works on similar systems such as the low-{\it z} luminous infrared galaxies (e.g., Eq. (1) in \citealt{bellocchi2013}). The dynamical masses we derive are (1.5\,$\pm$\,0.3)\,$\times$\,$10^9$\,$M_\odot$ and (8.5\,$\pm$\,2.4)\,$\times$\,$10^9$\,$M_\odot$ for galaxies E and W, respectively, assuming velocity dispersions of 103\,$\pm$\,17 km\,s$^{-1}$ and 132\,$\pm$\,19\,km\,s$^{-1}$ (obtained with the FWHM from our H$\alpha$ line), and the effective radii derived from our NIRCam images (see Sect. \ref{subSect:reff} for details). 
The total stellar masses are derived from the multiwavelength SED fitting and correspond to (15\,$\pm$\,9)\,$\times$\,$10^{8}$\,M$_{\odot}$ and (8\,$\pm$\,4)\,$\times$\,10$^{8}$\,M$_{\odot}$ for galaxies E and W. Thus, while there is a good agreement between the stellar and dynamical masses for galaxy E, an order of magnitude difference appears for galaxy W. The difference in galaxy W could be due to the presence of a dominant molecular gas component in this galaxy amounting to close to 90\% of the total mass. However, the stellar structure shows the presence of extended emission consisting of star-forming clumps along tidal tails (\citealt{sugahara_2024_rioja}). This suggests that this galaxy is far from relaxed, and therefore the estimated dynamical mass is highly uncertain.

\begin{figure} 
    \centering   \includegraphics[width=0.96\linewidth]{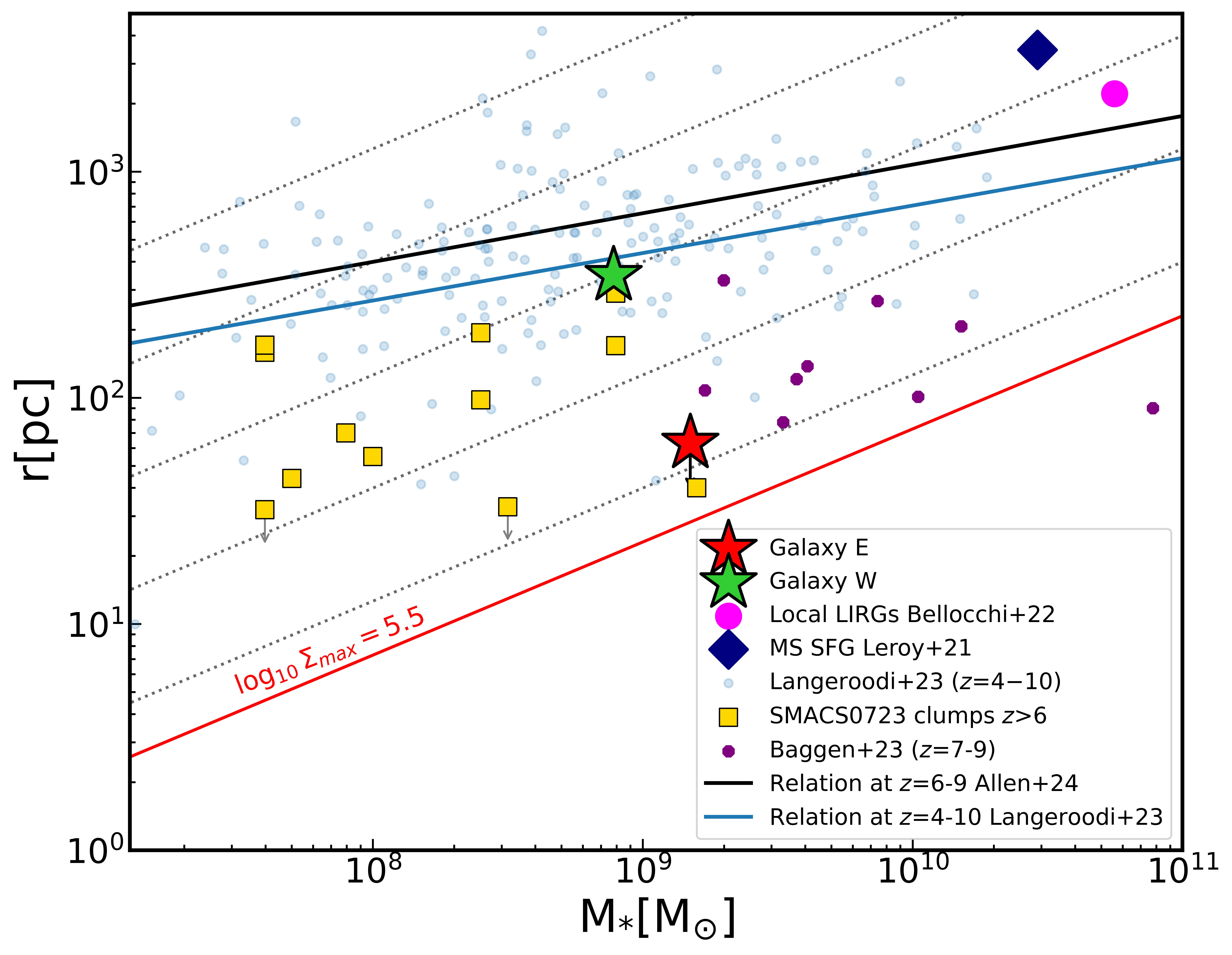}
    \caption{Mass-size relation for galaxies E (red star) and W (green star) in B14-65666,
    together with clumps of SMACS0723 at redshifts 6.4 to 8.5 (yellow squares, \citealt{Claeyssens_2023}) and a subsample of massive and compact galaxies from the CEERS program (purple circles, \citealt{baggen23}). The blue line represents the mass-size relation for galaxies (blue dots) at redshifts 4 to 10 
(\citealt{langeroodi_2023}), 
    while the solid black line indicates the mass-size relation for redshifts 6\,--\,9 derived for the NIRCam filter F150W (\citealt{allen2024galaxysizemassbuildup}). As B14-65666 is an interacting system, we also represent the mean values of the mass and size for local LIRGs and MS SFGs (\citealt{bellocchi_2022} and \citealt{leroy21}).
    Dotted gray lines represent constant stellar mass surface density for 10, 10$^2$, 10$^3$, 10$^4,$ and 10$^5$\,M$_\odot\, \text{pc}^{-2}$. The red line indicates the observed maximum value of stellar-mass surface density in clusters and nuclei of galaxies, 10$^{5.5}$\,M$_\odot\, \text{pc}^{-2}$.}
    \label{fig:mass-size}
\end{figure}

\subsection{Mass-size relation} \label{subSect:mass-size} 

The two galaxies of the B14-65666 system occupy a very different location in the stellar mass-size plane (see Fig. \ref{fig:mass-size}). 
Galaxy W, with an effective radius of 348 pc and a stellar mass of (8\,$\pm$\,4) $\times$ 10$^8$ M$_{\odot}$, is consistent with the expected values according to the relations derived for high-redshift galaxies from the JADES program (redshifts 4 to 10; \citealt{langeroodi_2023}) and from the CEERS, PRIMER-UDS, and PRIMER-COSMOS surveys (redshifts 3 to 9; \citealt{allen2024galaxysizemassbuildup}). Therefore, galaxy W appears as an average galaxy at approximately redshift 7. On the other hand, the second component of the system, galaxy E, appears to be deviating from the standard relations, showing a very small and compact size (upper limit of 63 pc to the effective radius) for a stellar mass of 15 ($\pm$\,9)\,$\times$\,10$^8$\,M$_{\odot}$. This translates into a high-stellar-mass surface density (6\,$\times$\,$10^{4}$\,M$_{\odot}$\,pc$^{-2}$ for E, versus 1\,$\times\,10^{3}$ M$_{\odot}$\,pc$^{-2}$ for W), close to the limit of the highest values ($10^{5.5}$\,M$_{\odot}$\,pc$^{-2}$) measured in the stellar clusters and nuclei of low-{\it z} galaxies (\citealt{grudic_19}). Galaxies as compact as galaxy E and with similar stellar surface densities are not uncommon at high-$z$ (\citealt{alvarez25}). Star-forming clumps and galaxies with an effective radius in the 30 to 200 pc range and a wide mass range (4\,$\times$\,10$^7$ to 2\,$\times$\,10$^{10}$ M$_{\odot}$) are already known at redshifts above 6 (see Fig. \ref{fig:mass-size}; \citealt{Claeyssens_2023}; \citealt{baggen23}).
Thus, B14-65666 is identified as a system that consists of a pair of interacting galaxies, with  similar stellar masses (within a factor of 2) but with one of the galaxies (E) being extremely compact and already showing stellar-mass surface densities close to those measured in the nuclei of galaxies at low redshifts. The galaxies of the B14-65666 system are much less massive ($\approx$\,$\times$\,20\,--\,30) and far more compact ($\approx$\,$\times$\,10\,--\,60) than the main-sequence and most luminous star-forming galaxies at low redshifts, with effective radii $\approx$\,3.5 kpc (\citealt{leroy21}) and $\approx$\,2.2 kpc (\citealt{bellocchi_2022}), respectively (see Fig. \ref{fig:mass-size}). Thus, galaxies E and W appear to be building up their stellar mass as they go through the advanced phases of the interaction/merging process.

\begin{figure} 
    \centering   \includegraphics[width=0.96\linewidth]{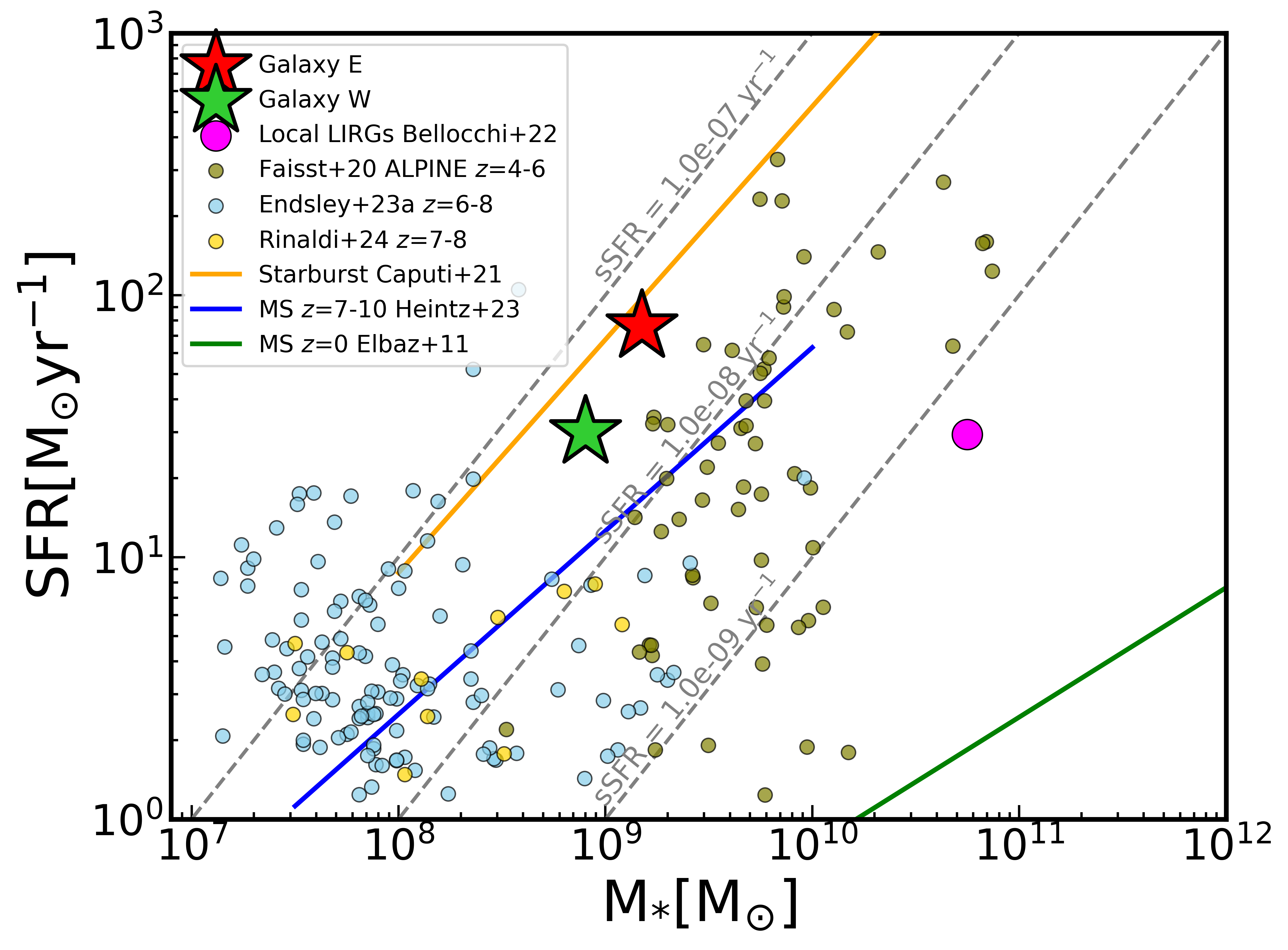}
    \caption{Stellar-mass recent SFR diagram. The large red star represents galaxy E, and the large green star represents galaxy W. We use the SFR traced by H$\alpha$ for solar metallicities. The orange line delimits the starburst area (\citealt{Caputi+17}). The blue line represents the MS for galaxies derived for redshift 7, and the green line shows the one for galaxies at redshift 0 (\citealt{heintz+23}, \citealt{elbaz_2011}). The dashed gray lines represent the lines of constant sSFR. We also present galaxies of redshifts 6\,--\,8 (\citealt{endsley_2023}) (blue dots), the main-sequence galaxies of ALPINE at redshifts 4.4\,--\,5.9 (\citealt{faisst_2020}, \citealt{fevre_2020}) (olive dots), strong H$\alpha$-emitters (HAEs) at $z$\,$\approx$\,7\,--\,8 (\citealt{rinaldi_24_halpha_emitters}) (yellow dots) and the mean for the LIRG sample (\citealt{bellocchi_2022}) (pink dot). 
   } 
    \label{fig:sfr-mass}
\end{figure}

\subsection{Starburst nature} \label{subSect:burstiness}

The analysis of the SFR derived with the H$\alpha$-versus-stellar-mass plane reveals that both galaxies E and W are well above the main sequence (MS) of galaxies at redshift 7 (\citealt{heintz+23}). Galaxies E and W of our system are situated between the region defined by \cite{Caputi+17} for H$\alpha$ emitters, the starburst (SB) region, and the MS (see Fig. \ref{fig:sfr-mass}). Such extremely compact starburst galaxies are expected to have short depletion times (<100 Myr, see \citealt{franco2020}), and therefore galaxy E may quickly quench out of gas exhaustion \---at least temporarily-- which is in line with a bursty/stochastic SFH. 

We compared the position in the SFR stellar-mass plot with more galaxies at similar redshifts (Fig. \ref{fig:sfr-mass}). The sSFRs of E and W (51 and 38 Gyr$^{-1}$, respectively) are similar to those of low-mass galaxies, but also to some very massive ones, such as REBELS-25 (sSFR=25 Gyr$^{-1}$), a relaxed, dynamically cold disc galaxy at a similar redshift, $z$\,=\,7.31 (\citealt{rowland24_rebels}). 
A recent large compilation of NIRSpec prism data and stacked spectra showed that the average galaxy becomes progressively more bursty with increasing redshift, as well as with increasing sSFR (\citealt{langeroodi24}). Based on this, B14-65666 seems slightly less bursty (but consistent within error bars) than the average galaxy at similar redshift and sSFR. UV-faint galaxies at redshifts from 6 to 8 from the Cosmic Evolution Early Release Science (CEERS) JWST/NIRCam  survey also show a bursty star formation history (\citealt{endsley_2023}). Galaxies from the ALMA-ALPINE [C\,II] survey at redshifts 4 to 6 (\citealt{fevre_2020}; \citealt{faisst_2020}) show  normal star formation following the MS of galaxies at these redshifts (\citealt{Speagle+14}). LIRGs at low redshifts, with typical stellar masses in the $10^{10}-10^{11}$\,M${_\odot}$ and SFR 30\,M${_\odot}$\,yr$^{-1}$ range (\citealt{bellocchi_2022}), are placed above the low-$z$ MS plane (\citealt{elbaz_2011}) and indicate that at high redshifts LIRGs are less massive but more compact.

\subsection{B14-65666: A merger of starburst galaxies in the early Universe} \label{subSect:simulations}

The picture that emerges from our analysis is that B14-65666 is a system of two galaxies in the process of interacting and merging. We identified one of the components (E) as an extremely compact source, characterized by a stellar-mass surface density close to that of the nuclei of low-{\it z} galaxies. On the other hand, the second component of the system (galaxy W) shows a more elongated structure with four fainter clumps (\citealt{sugahara_2024_rioja}), reminiscent of star-forming regions along the tidal tails produced in gas-rich interacting galaxies. A significant velocity offset of 175\,km\,s$^{-1}$ between the two galaxies supports the merger scenario. Furthermore, the estimated stellar masses for each component are on the order of $\sim$\,$10^9$\,M$_\odot$, which is three-to-four orders of magnitude higher than typical stellar masses of star-forming clumps in galactic disks ($10^5$\,--\,$10^6$\,M$_\odot$, \citealt{Adamo+24}; \citealt{mowla24}), reinforcing the interpretation that we are observing two galaxy-scale systems. The stellar-mass ratio between the components is $\sim$\,2:1, which is potentially consistent with a major merger. Additionally, we find differences in metallicity between the two components, with values derived using the SED-fitting of 0.24\,Z$_\odot$ for E and 0.31\,Z$_\odot$ for W; this is in agreement with line diagnostics from NIRSpec IFS (\citealt{gareth_2024}). Collectively, the presence of tidal features, velocity offset, distinct stellar populations and metallicities, and the high stellar masses all provide compelling evidence that B14-65666 is an example of a merging system of two galaxies. In addition, galaxies E and W are experiencing a bursty phase with an SFR well above that of the main-sequence star-forming galaxies of the same mass at those redshifts.

Recent simulations of the formation of galaxies during the EoR (\citealt{nazakoto_2024} and references therein) show that systems identified as clumpy --i.e., with two or more high-surface-density star-forming clumps as traced by emission lines-- represent 10$\%$ of all simulated sources. Moreover, most of the clumpy systems are produced by mergers, with the young stellar populations forming along the gas-rich tidal tails, and with specific SFR of the order of 50 Gyr$^{-1}$. However, these clumpy systems appear to be temporal and short-lived, merging into a single galaxy in about 50 Myr (\citealt{hashimoto23_A2744}). Under this scenario, the compact, and highly dense, galaxy E could already be a protobulge produced by the merger of previous clumps, while galaxy W would be in the process of merging of smaller and less massive clumps before the two main components, E and W, merge together in a single, more massive galaxy at a later time; i.e., at lower redshifts (see Figure 5 in \citealt{nazakoto_2024}). 

B14-65666 is not a unique system at those redshifts. Recent JWST observations show merging systems of star-forming galaxies such as the interacting system SPT0311-58, which resides in the core of an extremely massive protocluster at $z$\,=\,6.9 (\citealt{arribas_24_spt}); the source A2744-z7p9OD, the most distant protocluster at $z$\,=\,7.88 (\citealt{hashimoto23_A2744}); or the JWST’s Quintet, a merger of at least five galaxies containing over 17 clumps, at  $z$\,=\,6.7 (\citealt{hu2025circumgalacticenrichmentmultigalaxymerger}). The detection with the JWST and further studies of new systems such as these will allow us to understand the formation process of massive galaxies in the early Universe.

\section{Conclusions and summary}\label{Sec:conclusion_Summary}

We present mid-infrared imaging and spectroscopy of the $z$\,=\,7.15 Lyman-break galaxy B14-65666 taken with MIRI/JWST. The data provide the first detection of the H$\alpha$ emission line of this source, a system of two merging galaxies at the EoR. 
The study also combines the new MIRI data with NIRCam data (\citealt{sugahara_2024_rioja}) and archival high-angular-resolution ALMA imaging (\citealt{Hashimoto+18}) to derive additional physical properties. The main results are summarized below.

\begin{itemize}
\item The H$\alpha$-emitting gas shows a structure dominated by two spatially resolved sources (E and W) separated by a projected distance of 0.4 arcsec (i.e., 2.2\,kpc) and relative velocity of 175\,km\,s$^{-1}$.
The rest-frame UV light shows galaxy E as a very compact, unresolved source (upper limit for the effective radius of 63\,pc), while galaxy W is more extended (348\,pc) with the presence of fainter sources along an elongated structure.

{\item The SFR derived from the H$\alpha$ luminosity is 76 and 30  M$_{\odot}$\,yr$^{-1}$ (for solar metallicity as reference), for galaxies E and W, respectively. Other measurements of the SFR using hybrid tracers; H$\alpha$ and the luminosity TIR; and the UV-plus-TIR luminosities give values of 50 and 54 M$_{\odot}$\,yr$^{-1}$ and 66 and 88 M$_{\odot}$\,yr$^{-1}$ for galaxies E and W, respectively. These values are compatible within the uncertainties with the SFR derived from H$\alpha$. The ratio of H$\alpha$-to-UV-based SFR (burstiness) indicates that while galaxy E is in a mild bursty phase (ratio of 1.8), galaxy W is experiencing a more continuous burst (ratio of 0.8).}

\item The equivalent width of the H$\alpha$ line (832\,±\,100 and 536\,±\,78 $\AA$ for galaxies E and W, respectively) and the ionizing photon production efficiency ($\log(\zeta_\mathrm{ion})$\,=\,25.1\,±\,0.1 Hz\,erg$^{-1}$ and $\log(\zeta_\mathrm{ion})$\,=\,25.5\,±\,0.1 Hz\,erg$^{-1}$) in the two galaxies are within the range of values measured in other samples of galaxies at these redshifts. Their different locations in the production efficiency $-$ equivalent width plane, with galaxy W above and galaxy E below the relation for high-$z$ galaxies is interpreted as being due to slightly different ages in their young stellar populations.

\item The specific star formation rate (40\,--\,50\,Gyr$^{-1}$) indicates that the two galaxies involved in the B14-65666 system are experiencing an intense starburst phase, well above the typical values for galaxies of similar masses at those redshifts and similar to less massive ($\times$\,10) galaxies. 

\item The stellar-mass surface density of galaxy E is extreme ($>$\,6\,$\times$\,$10^{4}$\,M$_{\odot}$\,pc$^{-2}$) and close to the value measured in the nuclei of low-$z$ galaxies, while galaxy W has a surface density ($10^{3}$\,M$_{\odot}$\,pc$^{-2}$) similar to that of a high-$z$ galaxy.  

\item The H$\alpha$ emission-line profile shows a different gas kinematics, with galaxy W showing larger velocities (i.e., broader line) than galaxy E. The ionized gas kinematics traced by the H$\alpha$ line shows differences with respect to the neutral and highly ionized gas traced by the far-infrared [C\,II]\,158\,$\mu$m and [O\,III]\,88\,$\mu$m lines, respectively. This indicates a complex structure in the ISM, likely due to the internal substructure of the young and mature stellar populations and dust distribution in these galaxies on (sub)kiloparsec scales.
\end{itemize}
In summary, B14-65666 is a system consisting of two galaxies that are experiencing an extreme starburst phase due to the interaction/merging process. One of the galaxies (E) is already very compact and massive, with a stellar density close to that of the mature nucleus/bulge of low$-z$ galaxies, while the other one (W) is consistent with a more extended star formation, distributed in a few fainter clumps along a tidal tail.  These results agree with recent cosmological simulations of clumpy merging systems in the early Universe (\citealt{nazakoto_2024}). According to these simulations, this clumpy phase identified in B14-65666 is short-lived, with the bursty clumps merging on a short timescale (50 Myr) and forming a massive, single-nucleus galaxy at later times.  Detailed studies of systems similar to B14-65666 will help our understanding of the physical conditions and mechanisms involved in the formation process of massive galaxies and their evolution in the early Universe, just a few hundred Myr after the Big Bang and during the EoR.

\begin{acknowledgements}
We thank Michele Perna and Gareth Jones for discussions on B14-65666. CPJ, JAM, LC acknowledge support by grant PIB2021-127718NB-100 from the Spanish Ministry of Science and Innovation/State Agency of Research MCIN/AEI/10.13039/501100011033 and by “ERDF A way of making Europe”. ACG acknowledges support by JWST contract B0215/JWST-GO-02926. AB, JM and GÖ acknowledge support from the Swedish National Space Administration (SNSA). KIC acknowledges funding from the Dutch Research Council (NWO) through the award of the Vici Grant VI.C.212.036 and funding from the Netherlands Research School for Astronomy (NOVA). AAH acknowledges support from from grant PID2021-124665NB-I00 funded by the Spanish Ministry of Science and Innovation and the State Agency of Research MCIN/AEI/10.13039/501100011033 and ERDF A way of making Europe. This work was supported by research grants (VIL16599,VIL54489) from VILLUM FONDEN. JPP and TVT  acknowledge financial support from the UK Science and Technology Facilities Council, and the UK Space Agency. The project that gave rise to these results received the support of a fellowship from the “la Caixa” Foundation (ID 100010434). The fellowship code is LCF/BQ/PR24/12050015. PGPG and LC acknowledge support from grant PID2022-139567NB-I00 funded by Spanish Ministerio de Ciencia e Innovaci\'on MCIN/AEI/10.13039/501100011033, FEDER {\it Una manera de hacer Europa}.
EI acknowledges funding from the Netherlands Research School for Astronomy (NOVA). IJ acknowledge support from the Carlsberg Foundation (grant no CF20-0534) and the Cosmic Dawn Center is funded by the Danish National Research Foundation under grant No. 140.
SG acknowledges financial support from the Villum Young Investigator grant 37440 and 13160 and the Cosmic Dawn Center (DAWN), funded by the Danish National Research Foundation under grant DNRF140.
AE and FP acknowledge support through the German Space Agency DLR 50OS1501 and DLR 50OS2001 from 2015 to 2023.

The work presented is the effort of the entire MIRI team and the enthusiasm within the MIRI partnership is a significant factor in its success. MIRI draws on the scientific and technical expertise of the following organisations: Ames Research Center, USA; Airbus Defence and Space, UK; CEA-Irfu, Saclay, France; Centre Spatial de Liége, Belgium; Consejo Superior de Investigaciones Científicas, Spain; Carl Zeiss Optronics, Germany; Chalmers University of Technology, Sweden; Danish Space Research Institute, Denmark; Dublin Institute for Advanced Studies, Ireland; European Space Agency, Netherlands; ETCA, Belgium; ETH Zurich, Switzerland; Goddard Space Flight Center, USA; Institute d’Astrophysique Spatiale, France; Instituto Nacional de Técnica Aeroespacial, Spain; Institute for Astronomy, Edinburgh, UK; Jet Propulsion Laboratory, USA; Laboratoire d’Astrophysique de Marseille (LAM), France; Leiden University, Netherlands; Lockheed Advanced Technology Center (USA); NOVA Opt-IR group at Dwingeloo, Netherlands; Northrop Grumman, USA; Max-Planck Institut für Astronomie (MPIA), Heidelberg, Germany; Laboratoire d’Etudes Spatiales et d’Instrumentation en Astrophysique (LESIA), France; Paul Scherrer Institut, Switzerland; Raytheon Vision Systems, USA; RUAG Aerospace, Switzerland; Rutherford Appleton Laboratory (RAL Space), UK; Space Telescope Science Institute, USA; Toegepast- Natuurwetenschappelijk Onderzoek (TNO-TPD), Netherlands; UK Astronomy Technology Centre, UK; University College London, UK; University of Amsterdam, Netherlands; University of Arizona, USA; University of Cardiff, UK; University of Cologne, Germany; University of Ghent; University of Groningen, Netherlands; University of Leicester, UK; University of Leuven, Belgium; University of Stockholm, Sweden; Utah State University, USA. A portion of this work was carried out at the Jet Propulsion Laboratory, California Institute of Technology, under a contract with the National Aeronautics and Space Administration. We would like to thank the following National and International Funding Agencies for their support of the MIRI development: NASA; ESA; Belgian Science Policy Office; Centre Nationale D’Etudes Spatiales (CNES); Danish National Space Centre; Deutsches Zentrum fur Luft-und Raumfahrt (DLR); Enterprise Ireland; Ministerio De Economía y Competitividad; Netherlands Research School for Astronomy (NOVA); Netherlands Organisation for Scientific Research (NWO); Science and Technology Facilities Council; Swiss Space Office; Swedish National Space Board; UK Space Agency.

For the purpose of open access, the authors have applied a Creative Commons Attribution (CC BY) licence to the Author Accepted Manuscript version arising from this submission.

This work is based on observations made with the NASA/ESA/CSA James Webb Space Telescope. The data were obtained from the Mikulski Archive for Space Telescopes at the Space Telescope Science Institute, which is operated by the Association of Universities for Research in Astronomy (AURA), Inc., under NASA contract NAS 5-03127 for \textit{JWST}; and from the \href{https://jwst.esac.esa.int/archive/}{European \textit{JWST} archive (e\textit{JWST})} operated by the ESDC. 

This work has made use of data from the European Space Agency (ESA)
mission {\it Gaia} (\url{https://www.cosmos.esa.int/gaia}), processed by
the {\it Gaia} Data Processing and Analysis Consortium (DPAC,
\url{https://www.cosmos.esa.int/web/gaia/dpac/consortium}). Funding
for the DPAC has been provided by national institutions, in particular
the institutions participating in the {\it Gaia} Multilateral Agreement.

This research made use of Photutils, an Astropy package for detection and photometry of astronomical sources \citep{larry_bradley_2022_6825092}.

\end{acknowledgements}

\bibliographystyle{aa} 
\bibliography{bibliography} 

\end{document}